\documentclass[conference]{IEEEtran}
\ifCLASSINFOpdf
\else
\fi

\usepackage[T1]{fontenc}
\usepackage{booktabs}
\usepackage{amsmath}
\usepackage{amssymb}
\usepackage{enumitem}

\usepackage{multirow}

\usepackage{soul}
\usepackage{hyperref}
\hypersetup{
    colorlinks=true,
    linkcolor=black,
    filecolor=black, 
    citecolor=black,
    urlcolor=cyan,
}

\usepackage[framemethod=TikZ]{mdframed}

\usepackage[ruled, vlined, linesnumbered]{algorithm}
\usepackage[noend]{algpseudocode}
	\vspace{-0.8cm}%
\alglanguage{pseudocode}

\algnewcommand{\Initialize}[1]{%
  \State \textbf{Initialize:}
  \Statex \hspace*{\algorithmicindent}\parbox[t]{.8\linewidth}{\raggedright #1}
}

\usepackage{threeparttable}
\usepackage{hyperref}
\hypersetup{
    colorlinks=true,
    linkcolor=blue,
    filecolor=black,      
    urlcolor=cyan,
    pdfpagemode=FullScreen
    }
\usepackage{pifont}
\usepackage{listings}

\usepackage{url}
\urldef{\mailsa}\path|{alfred.hofmann, ursula.barth, ingrid.haas, frank.holzwarth,|
	\urldef{\mailsb}\path|anna.kramer, leonie.kunz, christine.reiss, nicole.sator,|
	\urldef{\mailsc}\path|erika.siebert-cole, peter.strasser, lncs}@springer.com|    
\usepackage[utf8]{inputenc}
\usepackage[english]{babel}

\usepackage{amsthm}

\usepackage{soul}

\usepackage{minibox}
\newcounter{observcntr}

\usepackage{minibox}
\newcounter{takeaway}
\newcommand*{\takeaway}[1]{%
    \stepcounter{takeaway}%
    \begin{center}
    \vspace{-2pt}
    \minibox[frame, rule=1pt,pad=3pt]{
        \begin{minipage}[t]{0.95\columnwidth}
        \textbf{Takeaway~\arabic{takeaway}:} \textit{#1}.
        \end{minipage}
    }
    \vspace{-2pt}
    \end{center}
}

\theoremstyle{definition}

\definecolor{R}{RGB}{0,0,150}

\usepackage{xspace}
\newcommand{\name}{\texttt{UniGuard}\xspace}

\theoremstyle{remark}

\newcommand{\eat}[1]{}

\begin{document}
\urlstyle{same}
%
\title{Kill Two Birds with One Stone! \\ Trajectory enabled Unified Online Detection of Adversarial Examples and Backdoor Attacks}

\author{
	\IEEEauthorblockN{Anmin Fu\IEEEauthorrefmark{2}, Fanyu Meng\IEEEauthorrefmark{2}, Huaibing Peng\IEEEauthorrefmark{2}, Hua Ma\IEEEauthorrefmark{3}, \\ Zhi Zhang\IEEEauthorrefmark{1}, Yifeng Zheng\IEEEauthorrefmark{4}, Willy Susilo\IEEEauthorrefmark{5},  Yansong Gao\IEEEauthorrefmark{1}\IEEEauthorrefmark{3} }

	\IEEEauthorblockA{\IEEEauthorrefmark{2} Nanjing University of Science and Technology, China.}
	
	\IEEEauthorblockA{\IEEEauthorrefmark{3} Data61, CSIRO, Australia.}
	\IEEEauthorblockA{\IEEEauthorrefmark{4} The Hong Kong Polytechnic University, China.}

	\IEEEauthorblockA{\IEEEauthorrefmark{5} University of Wollongong, Australia.}
	
	\IEEEauthorblockA{\IEEEauthorrefmark{1} The University of Western Australia, Australia.}
		
	\IEEEauthorblockA{Y. Gao is the Corresponding Author.}

}
	

%


\maketitle

\begin{abstract}
While adversarial example (AE) attacks and backdoor attacks are widely recognized as primary threats to the inference integrity of deep learning (DL) models, they are typically addressed separately. This fragmental approach is problematic, as an attacker can indeed exploit either threat to compromise the inference integrity of the backdoored model, without mandatorily relying on trigger-carrying samples, as AE is also available.
No unified detection framework currently exists to counter both, not to mention addressing each threat separately is already fraught with significant limitations.

Our innovative \name, is the first unified online detection framework capable of simultaneously addressing AE and backdoor attacks.
\name builds upon two key insights: first, both AE and backdoor attacks have to compromise the inference phase, making it possible to tackle them simultaneously during run-time via online detection. 
Second, an adversarial input—whether a perturbed sample in AE attacks or a trigger-carrying sample in backdoor attacks—exhibit distinctive trajectory signatures from a benign sample as it propagates through the layers of a DL model in forward inference. The propagation trajectory of the adversarial sample must deviate from that of its benign counterpart, otherwise, the adversarial objective cannot be fulfilled.
Detecting these trajectory signatures is inherently challenging due to their subtlety; \name overcomes this by treating the propagation trajectory as a time-series signal, leveraging LSTM and spectrum transformation to amplify differences between adversarial and benign trajectories that are subtle in the time domain. 
\name exceptional efficiency and effectiveness have been extensively validated across various modalities (image, text, and audio) and tasks (classification and regression) ranging from diverse model architectures (ResNet and foundational models of BERT) against a wide range of AE attacks (e.g., FGSM, PGD, BIM, CW, DeepFool, JSMA, Boundary Attack, PWWS, and TextBugger) and backdoor attacks, including challenging partial backdoors and dynamic triggers. 
When compared to state-of-the-art (SOTA) methods, including ContraNet (NDSS'22) specific for AE detection and TED (IEEE S\&P'24) specific for backdoor detection, \name consistently demonstrates superior performance, even when matched against each method's strengths in addressing their \textit{respective} threats---each SOTA \textit{fails to parts of attack strategies} while \name \textit{succeeds for all}.
\end{abstract}


%
\IEEEpeerreviewmaketitle

\section{Introduction}\label{sec:Intro}
There are three primary types of adversarial attacks~\cite{vassilev2024adversarial}: adversarial example (AE) attack~\cite{goodfellow2014explaining}, backdoor attack~\cite{gu2017badnets}, and poisoning attack~\cite{jagielski2018manipulating}. The AE attacks compromise the inference phase, poisoning attacks comprise the training phase~\cite{vassilev2024adversarial}\footnote{Note that backdoor attack can be achieved through poisoning attacks, but in this work we regard the poisoning attack as an availability type attack only according to~\cite{vassilev2024adversarial}, while backdoor attack focus on integrity breach.}, while backdoor attacks compromise both phases. Poisoning attacks indiscriminately affect all test samples, rendering them less stealthy as it is suspicious when the overall performance of the DL model experiences a notable decline. 
In contrast, AE attacks and backdoor attacks target only specific test samples while preserving the performance of unmanipulated test samples. This characteristic makes them stealthier and more difficult to detect by assessing validation performance on normal test samples. Given their dire consequence in industry~\cite{kumar2020adversarial}, great efforts have been made to counter them~\cite{ma2019nic,yang2022you,gao2019strip,mo2024robust,pal2024backdoor} (detailed in Section~\ref{sec:related}).

\vspace{0.1cm}
\noindent\textbf{Limitation of Exisiting Works.} However, it is noted that existing countermeasures are developed in two unrelated directions: they are specific to counter either AE attacks or backdoor attacks. This is apparently problematic because the attacker can indeed choose any one of them when launching an evasive attack post model deployment when the model is operated for providing inference. Thus, it is urgent to develop a defense that can counter them all at once, especially detecting them.

To the best of our knowledge, there is \textit{no unified detection} framework that can simultaneously detect both AE and backdoor attacks~\cite{goldblum2022dataset}, despite few works exploiting one type of attack (AE/backdoor) as a countermeasure for another type of attack (backdoor/AE)~\cite{wu2021adversarial,shan2020gotta,zhu2023ai,niu2024towards}. Conversely, one attack can also be used to strengthen the other attack~\cite{pang2020tale}. 
We hypothesize the absence of a unified detection framework lies in the fact that both attacks use different attack spaces with many diverse strategies. 
As for attacking surfaces, the backdoor is mainly introduced by manipulating the underlying model either through data poisoning or model poisoning or a hybrid of both, granting an attacker higher attacking capability and stealthiness. The attack can be performed through different backdoor type and trigger type designs (see Section~\ref{sec:relatedBD}), significantly challenging the detection, especially without controlling the training process but only the underlying model is accessible~\cite{peng2024model,mo2024robust}. In contrast to a backdoor attack, the AE does not need to tamper with the underling model. The AE only needs to compromise the inference phase by injecting imperceptible perturbations to make the underlying model misbehave. There are many AE attack strategies available to attackers ranging from white-box to black-box AE crafting methods. It is also challenging to counter AE attacks, especially without prior knowledge of attack strategies~\cite{bryniarski2021evading}.

Even when considering one specific attack threat—either an AE or a backdoor attack—existing detection methods struggle with limitations.
First, many detection methods are specifically devised for image modality---given AE~\cite{xu2017feature,wang2023addition,yang2022you,ho2022disco,xiang2022patchcleanser} and given backdoor attack~\cite{chou2020sentinet,pal2024backdoor,guo2023scale}
, which cannot be applied to other modalities such as text and audio. Second, these detections are overwhelmingly devised for classification tasks, which cannot be adopted for non-classifications tasks such as common regression tasks or it is unclear whether they are applicable~\cite{yang2022you,ma2019nic,zhu2023towards,chou2020sentinet,pal2024backdoor,guo2023scale,wang2023mm,mo2024robust}. Note that regression tasks, particularly time-series forecasting, have extensive applications across diverse industries such as finance, transportation, energy, and healthcare. Third, many detections are only effective against naively common attack types/methods, but not against other types, even though the attack can be chosen. For example, as shown in~\cite{mo2024robust}, many advanced online detection methods e.g., SCAn~\cite{tang2021demon} and Beatrix~\cite{ma2022beatrix} are ineffective against partial backdoor type combined with dynamic trigger type.

\vspace{0.1cm}
\noindent{\bf Detection Requirements.}
The above limitations necessitate the introduction of requirements for devising an effective detection method against AE and backdoor attacks at the same time:

\begin{enumerate}[noitemsep,topsep=0pt]
    \item \textit{RM1: Unified Framework.} 
    The detection framework should be unified. Specifically, it can identify both AEs and backdoor attacks concurrently, instead of dealing with them separately as \textit{all existing works fail to satisfy}.

    \item \textit{RM2: Modality Agnostic.} 
    The detection method should not only be specific to image modality but also adaptable to diverse data modalities, including text and audio.

    \item \textit{RM3: Task Agnostic.} 
    Despite existing detections being primarily devised for classification tasks, it is important that the detection can be extended to deal with non-classification tasks such as the common regression task.
    
    \item \textit{RM4: Attack Strategy Agnostic.} 
    The detection should be general to handle different attack strategies. For backdoor attacks,
    it should be robust against various trigger types and backdoor types. For AEs, it should be robust against numerous AE crafting methods e.g., either black-box or white-box.

    \item \textit{RM5: Run-time Overhead.} 
    As an online detection mechanism operating during the inference phase, minimizing runtime overhead is essential. The detection process should start parallel with the inference operation to ensure efficiency and minimize delays. Notably, many online detections including ContraNet and TED require waiting for the final inference results e.g., predicted label, as feedback/guidance before detection operation. 
\end{enumerate}

Now we have the following research question:
\begin{mdframed}[backgroundcolor=black!10,rightline=false,leftline=false,topline=false,bottomline=false,roundcorner=2mm]
    Is there a detection framework that can satisfy all five requirements?
\end{mdframed}

\vspace{0.1cm}
\noindent{\bf Our Solution.} We propose the \textit{first unified detection} framework \name that can essentially satisfy all these requirements. There are challenges confronted when satisfying all requirements. \name breaks these key challenges down into three components. 

\noindent$\blacksquare${\it Challenge 1: Addressing Requirements 1\& 2\& 3.} 
Despite its importance, addressing AE and backdoor attacks simultaneously poses greater challenges than addressing each separately, not to mention existing difficulties in tackling each attack. \name first examines their common vulnerable stage---both have to compromise the online inference phase. At this stage, adversarial inputs, whether carrying perturbations in AE attacks or triggers in backdoor attacks, aim to mislead the model's predictions. Therefore, \name is designed as an online detection per testing sample, sharing the advantage of fine-granularity~\cite{mo2024robust}.

To be modality agnostic, we should avoid the reliance on input space, such as by examining the image pixel characteristic used by~\cite{xu2017feature,guo2023scale,yang2022you}. The input space is data modality dependent. Similarly, to be task agnostic, we must avoid the dependence on the output, such as the predicted label used by~\cite{yang2022you,mo2024robust,tang2021demon,li2023ntd}. The output is inherently task-dependent, as each task produces distinct inference results tailored to its specific requirements. 

To this end, \name looks into the latent representation of the underlying model to obviate dependence on modality and task. Nonetheless, simply examining the latent representation of a given layer is not sufficient. Previous works including~\cite{tang2021demon,ma2022beatrix} have done so, but are ineffective even against the backdoor attack alone~\cite{mo2024robust}.

Instead, we leverage the consecutive latent representations across multiple layers. The \textcolor{blue}{\bf core insight} is that adversarial and benign inputs exhibit distinct propagation paths or trajectories during forward inference. In earlier network layers, adversarial samples closely mimic the representations of benign inputs to maintain semantic or visual similarity and avoid detection by human inspection. However, as the layers deepen, the representations of adversarial samples increasingly diverge to fulfill adversarial objectives. This phenomenon, termed propagation trajectory divergence, effectively highlights adversarial inputs. Take classification as an example, the latent representation of an adversarial must be aligned with the representation of its targeted class after a given layer, which deviates from the representation of its source class. 
This deviation can be either sharp or gradual. This is also the reason that detection upon a single layer representation only is insufficient because one cannot predetermine after which layer the transition occurs, but the trajectory can always reliably tell.

The challenge now lies in efficiently capturing the subtle propagation trajectory divergence. Adversarial and benign trajectories can significantly overlap, making their distinction difficult. To address this, \name constructively treats the propagation trajectory as a time-series signal and employs an LSTM network, known for its strength in capturing rich temporal information, to amplify the divergence. Furthermore, the temporal information is transformed into the spectral domain, also inspired by time-series signal processing, to further enhance the differentiation between adversarial and benign trajectories.

\noindent$\blacksquare${\it Challenge 2: Addressing Requirement 4.} A key obstacle in adversarial defense lies in the unpredictability of emerging or evolving attack strategies, compounded further when facing both AE and backdoor attacks simultaneously. 
Given the diverse array of strategies attackers might employ, designing a robust detection requires avoiding reliance on prior knowledge of specific attack strategies. To address this, we approach online adversarial sample detection as an \textit{anomaly detection} problem. Using the deep support vector data description (deep SVDD)~\cite{ruff2018deep}, trained exclusively on propagation trajectories of benign examples as a single-class meta-classifier, we effectively overcome this challenge.

\noindent$\blacksquare${\it Challenge 3: Addressing Requirement 5.} To ensure real-time detection, minimizing latency is essential for a runtime framework. \name addresses this by making core components (i.e, raw latent representation dimension reduction that is most time-consuming due to its high dimensionality) operate in parallel with the underlying model’s forward inference, maintaining a passive, non-intrusive design that functions seamlessly as a plug-in. Efficiency is also enhanced by reducing overhead through dual-layered dimension reduction. First, spatial dimension reduction is applied to raw latent representations using Uniform Manifold Approximation and Projection (UMAP)~\cite{healy2024uniform}. This operation runs in parallel with the inference process. Second, temporal dimension reduction is performed on the propagation trajectory via the LSTM-encoder. This operation can be performed once the last considered layer's latent representation reduction is completed. These reductions not only optimize runtime latency but also streamline \name's detection process, preserving its role as an efficient and almost parallel operation.

In summary,
\name innovatively leverages the propagation path or trajectory as the inherent characteristic for capturing adversarial inputs during run-time, regardless of adversarial inputs presenting AE perturbation or backdoor trigger.
This characteristic is agnostic to the data modality and task. To amplify the subtle propagation path, \name constructively treats the propagation path as a time-series signal to harness rich temporal information, which is further enhanced through spectrum domain transformation. To avoid any prior knowledge of a wide array of attacking strategies, the \name only leverages normal propagation trajectories to train an anomaly detector and treat all adversarial inputs as outliers. 

\vspace{0.10cm}
\noindent{\bf Contribution.} Our main contributions are fourfold\footnote{The source code of \name will be released upon publication.}.

\noindent$\bullet$ We propose \name, the first unified detection framework to address all inference phase related adversarial threats, either AE attack or backdoor attack simultaneously. This challenge has not been tackled by prior research.

\noindent$\bullet$ We leverage the inference propagation trajectory that is modality- and task-agnostic as an inherent characteristic to discern adversarial samples. The efficiency is promised by our innovative harness of temporal information encompassed by the trajectory through processing it as a time-series signal.

\noindent$\bullet$ We extensively validate and affirm the efficiency and effectiveness of \name across modality (image, text, audio) and task (classification and regression) on a wide range of AE attack strategies (white-box FGSM, PGD, BIM, CW, DeepFool, JSMA, and black-box Boundary Attack) and backdoor attack strategies (including challenging partial backdoor type and dynamic trigger type).

\noindent$\bullet$ We compare \name with SOTAs including ContraNet~\cite{yang2022you} (NDSS'22) as an AE detection SOTA and TED~\cite{mo2024robust} (IEEE S\&P'24) as a backdoor detection SOTA. We have also compared with NIC~\cite{ma2019nic} since it was shown that it can detect AEs and a very special backdoor, Trojan Attack~\cite{liu2018trojaning} sharing similarity with universal AE---our evaluations show it \textit{does not} work against other backdoors even the simplest BadNet\footnote{While we successfully reproduced ContraNet and TED using their released code, the NIC source code failed to function in our trials. Consequently, with great effort, we reimplemented NIC based on 
the version in~\cite{aldahdooh2022adversarial}.}.
\name demonstrates superior performance even by aligning with each SOTA's strength specific to either AE or backdoor threat. 


\section{Related Work}\label{sec:related}
\subsection{Adversarial Example Defenses}\label{sec:relatedAE}
To counter notorious AE attacks, three main approaches of model/training modification, input transformation and adversarial detection can be leveraged. 

\noindent\textbf{Model/Training Modification.} Adversarial training~\cite{rade2021reducing,madry2018towards,sehwag2021robust} is a widely adopted defense mechanism in this category, where (known) AEs are created and incorporated into the training process to enhance the model's robustness. This approach typically modifies the standard training procedure through regularization. However, it is inherently model-specific, necessitating re-training per DL model, which is increasingly costly as the complexity of the DL models grows. Some approaches modify the model architecture, such as introducing an outlier class~\cite{grosse2017statistical}, which requires re-training based on known AEs. This dependency on crafted AEs significantly increases computational overhead and leaves the defense susceptible to unseen AE attacks. Furthermore, architectural changes or the application of training regularizations frequently degrade the model's performance on benign examples, leading to a utility degradation~\cite{grosse2017statistical,rade2021reducing}.

\noindent\textbf{Input Transformation.} 
Input transformation can be performed through various techniques, such as JPEG compression~\cite{dziugaite2016study,liu2019feature}, bit reduction~\cite{guo2017countering,xu2017feature}, pixel deflection~\cite{prakash2018deflecting}, or random transformations~\cite{xie2017mitigating}. Beyond pixel-space defenses, AE images can be purified through generative models~\cite{samangouei2018defense,zhou2023eliminating} or diffusion models~\cite{nie2022diffusion,zhang2023diffsmooth}. However, these transformations inherently affect benign examples, leading to a trade-off in utility. Additionally, these methods are typically designed for specific imaging modalities, limiting their applicability across diverse domains. Moreover, they cannot detect AE attacks and trace their origins, thus deterring potential attackers.

\noindent\textbf{Adversarial Detection.} Leveraging neighborhood information of input examples, such as local intrinsic dimensionality~\cite{ma2018characterizing}, forms the basis for AE detection in various studies~\cite{papernot2018deep,feinman2017detecting,lee2018simple}. However, training associated detectors (meta-classifiers)~\cite{abusnaina2021adversarial,papernot2018deep} requires observing AEs, leaving them vulnerable to unseen attacks and incurring computational overhead due to neighborhood analysis~\cite{abusnaina2021adversarial}. Input preprocessing methods like Feature Squeezing~\cite{xu2017feature} and ADDITION~\cite{wang2023addition} detect AEs by analyzing prediction inconsistencies after transformations, but their effectiveness is limited against imperceptible perturbations. Certifiably AE detection~\cite{shumailov2020towards,xiang2023objectseeker} provides provable guarantees under bounded assumptions, such as perturbation magnitude. However, these methods are computationally expensive and easily bypassed when attackers violate assumptions, like perturbation constraints or specific norms. 

We consider NIC~\cite{ma2019nic} and ContraNet~\cite{yang2022you} for comparison. The NIC detects AEs by assessing label inconsistencies in intermediate layers, using internal classifiers for predictions. ContraNet~\cite{yang2022you} employs a class-conditional GAN to reconstruct images and compares semantic similarity for AE detection but sacrifices accuracy on benign samples~\cite{xu2017feature}. We compare \name with NIC as it has demonstrated effectiveness against a specific type of backdoor attack, namely Trojan Attack~\cite{liu2018trojaning}.
However, when we assess whether it generalizes to other backdoor attack types, we find that it does not. 
A potential reason is that the trigger used in the special Trojan attack shares similarities with universal adversarial perturbations. Since Trojan attacks first reverse-engineer a trigger based on the underlying model and then enhance it. We include ContraNet~\cite{yang2022you} (NDSS '2022) in our comparison as it represents a recent SOTA specific for detecting AEs.
\subsection{Backdoor Attack Defenses}\label{sec:relatedBD}

Generally, backdoors in machine learning models can be categorized into two main types: universal (source-class-agnostic) and partial (source-class-specific). 
A universal backdoor is independent of the input's source class, meaning that any input containing the trigger will activate the backdoor in an infected model. 
This enables the attacker to achieve the intended malicious behavior regardless of the input's original class. 
On the other hand, a partial backdoor~\cite{gao2019strip,wang2022cassock,tang2021demon,ma2022beatrix} is specific to an attack-chosen source class. 
For a partial backdoor to be activated, the input must both include the trigger and belong to the designated source class. 
If the input originates from a non-source class, the backdoor remains inactive, even when the trigger is present. 
Multiple-trigger-multiple-infected-classes~\cite{wang2019neural} and all-to-all backdoors~\cite{gu2017badnets} can be viewed as variants of these two primary types. It is important to distinguish between the trigger design/function \textsf{T}($\cdot$), which transforms an input $\textbf{x}$ into a trigger-carrying sample $\textsf{T}({\bf x})$, and the design of the backdoor mechanism itself, which dictates the behavior of the infected model~\cite{ma2024hcb}. 
Research on trigger designs—such as patch triggers in BadNet~\cite{gu2017badnets}, blending triggers~\cite{chen2017targeted}, and imperceptible triggers in WaNet~\cite{nguyen2021wanet} and ISSBA~\cite{li2021invisible}—focuses on the transformation $\textsf{T}({\bf x})$ applied to the input. 
Notably, the same trigger type can facilitate different backdoor types, emphasizing that backdoor type and trigger type are distinct and orthogonal concepts.

For backdoor defenses, they can be classified into three categories: prevention/removal, model-based detection, and data-based detection. 

\noindent{\bf Prevention/Removal.} Preventive and removable defenses can be customized for various scenarios depending on the defender's capabilities. 
When the defender has full control over the training process, it is possible to train a clean model even on a poisoned dataset~\cite{li2021anti,tao2022model}. 
This can be achieved through delicately designed training procedures~\cite{tao2022model,huang2021backdoor}, knowledge distillation~\cite{gong2023redeem}, or selective amnesia techniques~\cite{zhu2023selective}. 
For pre-deployed models, backdoor effects can be mitigated via fine-pruning~\cite{liu2018fine} or model orthogonalization~\cite{tao2022model}. 
Additionally, deployed models can address triggers dynamically by purifying incoming inputs~\cite{shi2023black}. 
These defenses are blindly applied to any dataset or model regardless of whether they are indeed infected. 
However, in practical settings, datasets or models are often benign. 
Blindly applying such defenses can result in high computational overhead~\cite{tao2022model} or even degrade model performance, posing challenges for real-world adoption.

\vspace{0.2cm}
\noindent{\bf Model-based Detection.} 
This category focuses on identifying whether a given model is backdoored. 
If an infection is detected, the backdoor trigger can be reverse-engineered, and unlearning techniques can be applied to eliminate its effects. 
Unlearning approaches, such as Neural Cleanse~\cite{wang2019neural}, or other methods~\cite{dong2021black,guo2020towards} based on similar principles, achieve this by neutralizing the backdoor behavior. 
ABS~\cite{liu2019abs} reverse-engineers potential triggers by assuming that only a small number of neurons are compromised. 
Similarly, DeepInspect~\cite{chen2019deepinspect} employs AI-against-AI strategies, leveraging a GAN to reconstruct triggers. 

Others are only to justify whether the model is compromised or not but are not to reverse-engineer the trigger like above defenses do. AI-against-AI approaches, such as training meta-classifiers to evaluate models-under-test, have also been explored in methods like MNTD~\cite{xu2021detecting} and ULP~\cite{kolouri2020universal}. It should be noted that these AI-against-AI approaches are sensitive to hyperparameters used for the meta-model training, which exhibit low robustness in addition to their high computational overhead. Other detection techniques analyze latent representations using statistical methods, including Beatrix~\cite{ma2022beatrix}, Trojan Signature~\cite{fields2021trojan}, and MM-BD~\cite{wang2023mm}--- the latter two are lightweight.

\noindent{\bf Data-based Detection.} 
It can be classified into offline and online approaches. Offline methods, which typically have full access to the training dataset, include Beatrix~\cite{ma2022beatrix}, which examines higher-order latent representations to distinguish benign from poisoned samples. Beatrix requires reserving a small clean dataset. In contrast, TellTale~\cite{gao2025try} analyzes the training trajectory rather than static spatial representations, effectively identifying poisoned samples without relying on a clean reserved dataset.
Unlike these reactive approaches~\cite{ma2022beatrix,gao2025try}, proactive methods like CT~\cite{qi2023towards} and ASSET~\cite{pan2023asset} actively amplify the differences between clean and poisoned samples, facilitating detection.

Online detection techniques leverage properties such as saliency maps in trigger-stamped regions~\cite{chou2020sentinet}, strong confidence against perturbations~\cite{gao2019strip}, and topology evolution dynamics~\cite{mo2024robust}. Beatrix~\cite{ma2022beatrix} can also be applied for online detection but requires the prior identification of dozens of trigger-carrying samples, which is problematic since such attempts already reaching a successful attack. Online detection methods~\cite{gao2019strip,chou2020sentinet,pal2024backdoor,guo2023scale} are often constrained by data modality, being more effective~\cite{gao2019strip} or even exclusively applicable to image modalities~\cite{chou2020sentinet,pal2024backdoor,guo2023scale}. Additionally, they struggle to handle advanced backdoor and trigger designs~\cite{chou2020sentinet,gao2019strip,pal2024backdoor}. TED~\cite{mo2024robust}, a SOTA approach, addresses these challenges, including modality generalization and independence from trigger and backdoor types. Thus, we compare \name with TED for online per-sample detection of backdoor attacks. However, TED is limited to classification tasks, whereas \name extends to non-classification tasks, such as regression, which we validate in our experiments.

\section{\name}
We first define the threat model and then overview \name, followed by detailed implementations of its core components.
\subsection{Threat Model}
Our threat model is aligned with AE online detection~\cite{ma2019nic,yang2022you} and trigger-carrying sample online backdoor detection~\cite{gao2019strip,mo2024robust} with which we also compare~\cite{ma2019nic,yang2022you,mo2024robust}.

\noindent\textbf{Defender Goal.}
The primary objective of the defender is to develop an effective input-level adversarial detection strategy during inference. For backdoor attacks, this involves determining whether a testing sample presents a trigger. For AE attacks, it detects whether a testing sample is manipulated to mislead the underlying model. Notably, \name has to detect AEs \textit{regardless of whether the underlying model is backdoored}. Since benign models tend to be more common in practice, \name must perform effectively in such scenarios. In addition, \name should exhibit low false positive rates for benign samples. Put them together, the \name treats these adversarial testing samples regardless of whether trigger-carrying samples are in a backdoor or AE samples as anomalies that are different from benign samples. 


\noindent\textbf{Defender Capability and Knowledge.}
The defender is assumed to have full access to the given model (i.e., all intermediate layers) but without the ability to interfere with the model's training process or knowledge of the training procedure. Additionally, the defender has access to a small subset of benign samples for constructing the detector offline, which is mandatory for all existing online adversarial sample-level detection methods~\cite{ma2019nic,yang2022you,gao2019strip,mo2024robust}.
This assumption is reasonable, as the defender can always reserve a small portion of normal testing samples. As we show in Section~\ref{sec:number}, reserving only 100 samples (0.2\% rate of the training dataset of CIFAR10) is satisfactory.
However, the defender operates without prior knowledge of attack strategies, including the type of attack deployed or whether the model has been backdoored. 

\subsection{Overview}
\begin{figure*}[h]
	\centering
	\includegraphics[trim=0 0 0 0,clip,width=0.95 \textwidth]{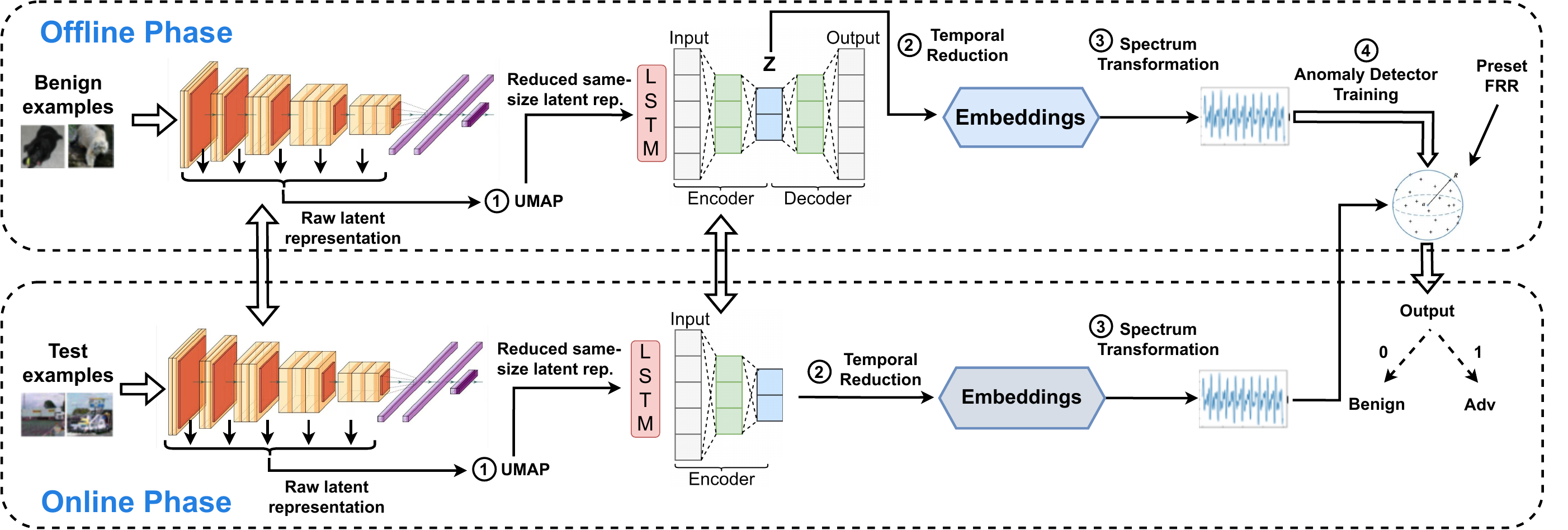}
	\caption{\name overview.}
	\label{fig:uniguard_overview}
\end{figure*}

As illustrated in Figure~\ref{fig:uniguard_overview},
\name consists of two phases: offline phase and online phase. During the offline phase, it constructs a detector (in particular, Deep-SVDD~\cite{ruff2018deep,tax2004support}) without requiring any prior knowledge of attack strategies. In the online phase, this detector is employed to classify incoming inputs or test samples, determining whether they are adversarial. Adversarial samples identified by \name can originate from either AEs or backdoor attacks. 

\vspace{0.1cm}
\noindent$\blacksquare$\textbf{Offline Phase:} This phase has a sequence of steps aimed at constructing a robust detector. During this phase, the underlying model performs inference on a small set of reserved benign samples:  

\begin{itemize}[leftmargin=*]

    \item \textit{Layer-Dimensionality Reduction}: In Step \textcircled{\small 1}, the raw latent representation of each sample, generated by every layer of the model, is reduced to a low-dimensional vector with the same dimension. This ensures a uniform dimensionality for subsequent processing. Note the size of raw latent representation per layer can vary significantly.

    \item \textit{Temporal Reduction}: In Step \textcircled{\small 2}, the layer-wise reduced vectors from Step \textcircled{\small 1} are fed into an LSTM-based encoder-decoder. It is to perform temporal-space reduction by treating each reduced latent presentation vector per layer as sampled per time point of the LSTM input, producing a compact representation denoted as \textbf{z} that captures rich temporal-like information.  

    \item \textit{Spectrum Transformation}: In Step \textcircled{\small 3}, the compact temporal representations \textbf{z} is transformed into the spectrum domain, enabling more effective anomaly detection.  

    \item \textit{Anomaly Detector Training}: In Step \textcircled{\small 4}, the spectrums derived from benign samples are used to train Deep-SVDD, an anomaly detector designed to identify adversarial inputs. Adversarial inputs, whether adversarial examples or trigger-carrying samples, are uniformly treated as anomalies. This universal treatment enables \name to function effectively without requiring prior knowledge of specific attack strategies or their threat origins (i.e., AE or backdoor).
\end{itemize}


\vspace{0.1cm}
\noindent$\blacksquare$\textbf{Online Phase:} This phase is straightforward. Given an incoming/testing sample, it goes through \textcircled{\small 1} to \textcircled{\small 3} to obtain the corresponding spectrum, which is then fed into the Deep-SVDD. The Deep-SVDD determines whether it is benign or adversarial.

\subsection{Implementation}\label{sec:implementation} 
Essentially, each step corresponds to a key component design of \name, which implementation is detailed below.

\subsubsection{Layer-Dimensionality Reduction} We feed each benign sample into the model and extract the activations of each convolutional layer as its raw latent representation (note similar operation can be applied to different model architectures, as for BERT model validated in Section~\ref{sec:general}). Note that we do not extract representations from other layers, such as pooling layers or normalization layers, as convolutional layers are dominant layers. 

However, the direct usage of raw latent representations face two challenges: high dimensionality and inconsistent size across layers. For example, in ResNet18, the latent representation size of the first convolutional layer [1, 64, 32, 32] differs significantly from that of the last convolutional layer  [1, 512, 4, 4]. High-dimensional metadata imposes heavy computational overhead, and size inconsistency complicates unified post-processing, such as using the representations as input for the LSTM-encoder for temporal reduction in subsequent steps.

To address these challenges, we apply dimensionality reduction to achieve the same reduced size, e.g., $1\times 400$. While PCA is a common choice for such tasks, we utilize UMAP~\cite{mcinnes2018umap}, which is significantly more efficient than PCA. Detailed comparisons of computational overhead between PCA and UMAP are provided in Appendix~\ref{sec:PcaVsUmap}.

\subsubsection{Temporal Reduction} The sample forward inference is performed sequentially, layer by layer, along the network's depth. We treat this sequentially layer-wise propagated signal as a time-series signal, which is expected to carry rich temporal information. To leverage this temporal information, often overlooked in previous works, we propose the use of a lightweight long short-term memory (LSTM) network, which excels at capturing temporal dependencies.

Specifically, an LSTM-based encoder-decoder is employed to further reduce the dimensionality of the reduced latent representation vectors obtained from the previous step. Each latent vector, treated as a signal sampled at a given time step, is fed sequentially into the LSTM encoder-decoder during its training. After training, the paired decoder is discarded, and the bottleneck representation \textbf{z}, located at the junction of the encoder-decoder, is extracted as the temporally reduced representation. In fact, this step extracts the most important features of the information, i.e., it improves the signal-to-noise ratio (SNR). For example, in the case of ResNet18, there are 18 latent vectors of size \( 1 \times 400 \). After LSTM-based encoding, the extracted temporal representation is reduced to a size of e.g., \( 1 \times 60 \), as used in our experiments.
More specifically, we employ a two-layer bidirectional LSTM model structured as an autoencoder for noise reduction and dimensionality compression. To mitigate overfitting, a dropout rate of 0.2 is applied to both the encoder and decoder. The training process spans 100 epochs, with the autoencoder typically requiring only a few minutes to train. This efficiency justifies our choice over more complex models, such as BERT, which can also capture temporal information but entail significantly higher computational costs.

\subsubsection{Spectrum Transformation.} 
In signal processing, analyzing signals in the frequency domain often uncovers features that are less apparent in the time domain. Building on our earlier analogy, where the model's layer-axis latent vector is treated as a time-series signal, applying a spectrum transformation aligns well with its sequential nature. Consequently, we leverage the Fast Fourier Transform (FFT) to convert the temporal latent reduced vector \textbf{z} into its spectrum representation. This transformation further enhances the contrast between anomalous and benign samples, enabling more reliable differentiation.

\subsubsection{Anomaly Detector Training} 
Deep SVDD~\cite{ruff2018deep} extends the classical Support Vector Data Description algorithm to high-dimensional data by integrating deep neural networks. Its primary objective is to learn a compact representation of normal data within a low-dimensional embedding space, often modeled as a hypersphere, and minimize the hypersphere's volume. This approach ensures that the most fundamental features of normal data are captured, allowing anomalies to be identified as instances lying outside the learned boundary. 

In our experiments, we employ spectrum transformations derived from benign samples to train the Deep SVDD detector. Notably, Deep SVDD incorporates a preset FRR as a critical hyperparameter for its operation\footnote{\url{https://github.com/yzhao062/pyod/blob/master/examples/deepsvdd_example.py}}. During Deep SVDD training, the defender can configure a small FRR (typically ranging from 1\% to 5\% in real-world applications). The FRR represents the likelihood of a benign sample being misclassified as anomalous, resulting in its rejection. Once trained, the Deep SVDD detector utilizes thresholds aligned with the preset FRR to differentiate between benign and anomalous samples effectively.

\section{Experiments}\label{sec:experiment}
This section extensively evaluates \name on the image dataset. We compare \name with SOTAs and also affirm its scalability with complicated model architecture or datasets.


\subsection{Setup} \label{sec:setting}
\subsubsection{Dataset and Model}
For the sake of extensive evaluation purposes and end-to-end comparisons with~\cite{ma2019nic,yang2022you,mo2024robust}, we mainly use the CIFAR10 dataset and ResNet18 model architecture, which are also extensively used by SOTAs~\cite{ma2019nic,mo2024robust}. The scalability evaluation on deeper networks, such as ResNet152, and more complex datasets, like Tiny-ImageNet, is detailed in Section~\ref{sec:scalability}.

\subsubsection{Attacks} \name counters backdoor attacks by detecting trigger-carrying samples and AE attacks by detecting the perturbed adversarial examples.

\begin{figure}[h]
    \centering
    \includegraphics[trim=0 0 0 0,clip,width=1.0\linewidth]{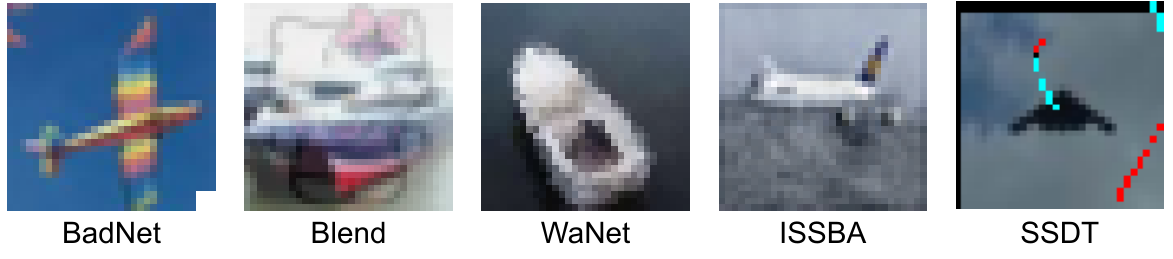}
    \caption{Examples of five different trigger carrying samples (CIFAR10). The WaNet and ISSBA triggers are imperceptible.
    }
    \label{fig:backdoor_compare}
\end{figure}

\noindent$\bullet$ For backdoor attacks, the source-specific and dynamic trigger (SSDT) attack is a newly crafted attack~\cite{mo2024robust} that combines the strength of partial backdoor type and dynamic trigger type design. It successfully evades all previous online sample-level backdoor detection methods except the SOTA of TED (IEEE S\&P '2024)~\cite{mo2024robust}. Therefore, we take SSDT as the most evasive backdoor for our evaluation. In addition, the other four trigger designs of BadNet~\cite{gu2017badnets}, WaNet~\cite{nguyen2021wanet}, ISSBA~\cite{li2021invisible}, and Blend~\cite{chen2017targeted} are considered because they have been used in SOTA~\cite{mo2024robust}. For these four trigger types, the universal backdoor is implanted, which means any sample carrying the trigger will fire the backdoor and force the infected model to exhibit a preset backdoor pattern. 

BadNet uses a white-square patch located at a corner as the trigger. WaNet distorts the global structure of images to craft trigger samples. Blend adds a transparent image on the poisoned image as the trigger. Here, the image of Hello Kitty is used with a transparency of $\alpha$ = 0.1. For each trigger type, a trigger-carrying example of CIFAR10 is exemplified in Figure ~\ref{fig:backdoor_compare}. Except for BadNet, all other triggers are designed to be imperceptible, so they are unsuspicious to humans when the trigger-carrying sample is used for launching online attacks.

The poisoning rate is set to 1$\%$ for all triggers except WaNet. A low poisoning rate is preferable in practice, as it enables the attacker to remain stealthy while minimizing the attack budget, provided the attack success rate (ASR) remains sufficient. WaNet, however, cannot achieve a high ASR with a low poisoning rate, such as 1$\%$, and thus requires a poisoning rate of 5\%.

\noindent$\bullet$ To evaluate \name against AE attacks, we test it on seven widely recognized AE attack methods: six white-box attacks and one black-box attack. The white-box attacks include the Fast Gradient Sign Method (FGSM)~\cite{goodfellow2014explaining}, Projected Gradient Descent (PGD)~\cite{madry2017towards}, Basic Iterative Method (BIM)~\cite{kurakin2018adversarial}, Carlini and Wagner Attack (CW)~\cite{carlini2017towards}, DeepFool~\cite{moosavi2016deepfool}, and the Jacobian Saliency Map Attack (JSMA)~\cite{papernot2016limitations}. The black-box attack we evaluate is the Boundary Attack~\cite{brendel2017decision}.

All these attack methods impose constraints on perturbations using \(\ell\)-norm metrics such as \(\ell_0\), \(\ell_1\), \(\ell_2\), and \(\ell_\infty\)~\cite{machado2021adversarial} for the sake of restricting the perturbation to be imperceptible that is one of most desired characteristics of AE. In our experiments, we configure the \(\ell_2\)-norm for DeepFool, \(\ell_0\)-norm for JSMA, and \(\ell_\infty\)-norm for FGSM, BIM, PGD, CW, and Boundary attacks. For attacks like FGSM, BIM, and PGD, the perturbation magnitude \(\epsilon\) is adjustable; we set \(\epsilon = 8/255\) to ensure that AEs retain semantic similarity to benign examples. For the remaining AE attacks, \(\epsilon\) is typically fixed by default.

\subsection{Metric} Four metrics are mainly used in the evaluations.

\noindent$\bullet$ {\bf Clean Data Accuracy (CDA)}: CDA measures the accuracy of the backdoored model on normal test samples in the absence of the trigger. The CDA of the backdoored model should closely match that of its clean model counterpart.

\noindent$\bullet$ {\bf Attack Success Rate (ASR)}: ASR evaluates the accuracy of the backdoored model on trigger-carrying samples. The ASR should ideally be as high as possible, typically reaching 100\% or close to it, especially for universal backdoor type.

\noindent$\bullet$ {\bf False Rejection Rate (FRR)}: FRR quantifies the probability of a benign sample being incorrectly flagged by \name as an adversarial sample. Adversarial samples are either adversarial examples or trigger-carrying samples. The FRR should be minimized, ideally achieving 0\%.

\noindent$\bullet$ {\bf Detection Accuracy}: Detection accuracy measures the likelihood of an adversarial sample being correctly identified as adversarial by \name. Ideally, the detection accuracy should reach 100\%.

\subsection{Results}
\subsubsection{Backdoor Attack} For BadNet, WaNet, ISSBA, Blend based attacks, 1,000 trigger-carrying samples are created for evaluation. Meanwhile, 3,000 benign samples are tested. As for the SSDT attack, 1,000 trigger-carrying samples are used because there are only 1,000 testing examples in the CIFAR10 dataset for a given class. More specifically, for SSDT, it is a partial backdoor attack where the source class is from only a single class. Meanwhile, 1,000 benign samples are tested for the sake of balance. 

\begin{table}
\centering 
\caption{\name performance against backdoor attacks. 
}
\resizebox{0.35 \textwidth}{!}
{
\begin{tabular}{cccccc}
\toprule
\multirow{2}{*}{\begin{tabular}[c]{@{}c@{}}Preset\\ FRR(\%) \end{tabular}} &
\multicolumn{5}{c}{Trigger type  (CIFAR10 and ResNet18)} \\ 
\cmidrule(l){2-6}
 &BadNet  &  WaNet &  ISSBA & Blend & SSDT \\
\midrule
\multirow{1}*{1} 
& 99.18\% & 99.26\% & 98.74\% & 99.39\% & 99.02\%\\
\multirow{1}*{3} 
& 99.26\% & 99.33\% & 98.88\% & 99.47\% & 99.38\% \\
\multirow{1}*{5} 
& 99.68\% & 99.59\% & 99.44\% & 99.89\% & 99.55\% \\
\bottomrule
\end{tabular}
}
\label{tab:BD_performance}
\end{table}

\begin{table}
\centering 
\caption{\name online FRR of backdoor detection.
}
\resizebox{0.32 \textwidth}{!}
{
\begin{tabular}{cccccc}
\toprule
\multirow{2}{*}{\begin{tabular}[c]{@{}c@{}}Preset\\ FRR(\%) \end{tabular}} &
\multicolumn{5}{c}{Trigger type (CIFAR10 and ResNet18)} \\ 
\cmidrule(l){2-6}
 &BadNet  &  WaNet &  ISSBA & Blend & SSDT \\
\midrule
\multirow{1}*{1} 
& 0.58\% & 0.82\% & 1.00\% & 0.60\% & 1.04\%\\
\multirow{1}*{3} 
& 2.42\% & 2.70\% & 2.04\% & 2.42\% & 2.08\% \\
\multirow{1}*{5} 
& 3.56\% & 4.66\% & 5.00\% & 3.66\% & 4.72\% \\
\bottomrule
\end{tabular}
}
\label{tab:BD_online_FRR}
\end{table}

The backdoor detection performance of \name is summarized in Table~\ref{tab:BD_performance}. The results highlight its consistently superior performance, which remains robust regardless of the backdoor type or trigger type design. More precisely, SSDT represents an advanced partial backdoor attack, while ISSBA and SSDT utilize dynamic and imperceptible triggers that vary for each sample. Despite these complexities, \name achieves no less than 99\% detection accuracy for trigger-carrying samples across all attack types.  

This performance holds even with a very low preset FRR of 1\%, which means \name to falsely classify only 1\% of benign samples as adversarial, striking a practical balance. Additionally, the observed online FRR closely aligns with the preset value, validating the practicality of using preset FRR in the DeepSVD anomaly detector to predetermine a tolerable FRR for the \name framework. Specifically, we have extensively evaluated the online FRR and shown detailed results in Table~\ref{tab:BD_online_FRR}. The online FRR under any backdoor/trigger attack is very close to the preset FRR.  

Increasing the tolerance for FRR (from 1\% to 5\%) further improves detection accuracy as expected (from 99.02\% to 99.53\% for SSDT). However, given that \name already delivers exceptionally high detection accuracy ($>$99\%) with a 1\% FRR, adopting the smaller FRR is more advantageous. Higher FRR values, such as 5\%, yield only marginal improvements in detection accuracy, making the 1\% FRR a more practical and efficient choice.

\takeaway{
\name is highly effective in detecting trigger-carrying samples regardless of their dependent backdoor types including the challenging partial backdoor used in SSDT) and trigger types including imperceptible and dynamic triggers
}

\subsubsection{Adversarial Example}
To demonstrate that \name is universal to both backdoor attacks and AEs, we evaluate \name performance against AEs on \textit{backdoored models} (\name performance against AEs on clean models is detailed in Section~\ref{sec:benignModel}). Those infected models are trained with the five trigger types mentioned above, namely BadNet, WaNet, ISSBA, Blend, and SSDT, respectively, as the underlying models. That means, for each of the five backdoored models, we evaluate \textit{all} AE methods. We use the Adversarial Robustness Toolbox\footnote{\url{https://github.com/Trusted-AI/adversarial-robustness-toolbox}} to implement these AE attacks. 
For all six white-box AE attacks, we use 3000 benign examples to craft AEs, the number of successful AEs depends on the attack success rate of each type of AE attack, e.g., varying from 77\% to 96\%. Note that the AEs we evaluated were those that can achieve the AE attack objective successfully.
Since implementing black-box AE attacks (e.g., boundary attacks), consumes a lot of time, we select only 200 samples in the test dataset (which takes about 10 hours) to craft AEs. The success rate of the boundary attack is 72\%, i.e., 144 AEs were successfully crafted.

\begin{table}[h]
\centering 
\caption{\name detection performance against AEs when the model is backdoored (CIFAR10 and ResNet18).
}
\resizebox{0.85\linewidth}{!}
{

\begin{tabular}{ccccccc}
\toprule
\multirow{2}{*}{\begin{tabular}[c]{@{}c@{}}AE\\ Strategy \end{tabular}} & \multirow{2}{*}{\begin{tabular}[c]{@{}c@{}}Preset\\ FRR(\%) \end{tabular}} & \multicolumn{5}{c}{Backdoored Model}      \\  \cmidrule(l){3-7}
&      & BadNet & WaNet & ISSBA & Blend & SSDT \\ \midrule
\multirow{3}{*}{\begin{tabular}[c]{@{}c@{}}FGSM\\ (\(\ell_\infty \)-norm)\end{tabular}}      
& 1    & 97.45\%      & 98.07\%     & 98.54\%     & 98.65\%     & 97.02\%    \\
& 3    & 97.72\%      & 98.61\%     & 99.31\%     & 99.39\%     & 97.35\%    \\
& 5    & 99.05\%      & 99.20\%     & 99.75\%     & 99.86\%     & 99.12\%    \\ \midrule
\multirow{3}{*}{\begin{tabular}[c]{@{}c@{}}BIM\\ (\(\ell_\infty \)-norm)\end{tabular} }       
& 1    & 96.37\%      & 96.77\%     & 97.87\%     & 97.44\%     & 96.41\%    \\
& 3    & 97.15\%      & 97.05\%     & 98.35\%     & 97.95\%     & 97.16\%    \\
& 5    & 99.34\%      & 97.59\%     & 99.43\%     & 98.48\%     & 98.97\%    \\ \midrule
\multirow{3}{*}{\begin{tabular}[c]{@{}c@{}}PGD\\ (\(\ell_\infty \)-norm)\end{tabular}}       
& 1    & 96.56\%      & 97.98\%     & 97.71\%     & 96.25\%     & 96.33\%    \\
& 3    & 97.30\%      & 98.62\%     & 98.23\%     & 97.53\%     & 97.17\%    \\
& 5    & 99.72\%      & 99.34\%     & 98.66\%     & 98.10\%     & 99.01\%    \\ \midrule
\multirow{3}{*}{\begin{tabular}[c]{@{}c@{}}CW\\ (\(\ell_\infty \)-norm)\end{tabular}}        
& 1    & 96.55\%      & 97.32\%     & 97.27\%     & 96.00\%     & 97.10\%    \\
& 3    & 97.42\%      & 97.39\%     & 98.04\%     & 96.35\%     & 97.86\%    \\
& 5    & 99.35\%      & 98.20\%     & 98.70\%     & 97.16\%     & 98.43\%    \\ \midrule
\multirow{3}{*}{\begin{tabular}[c]{@{}c@{}}JSMA\\ (\(\ell_0 \)-norm)\end{tabular}}      
& 1    & 96.75\%      & 98.61\%     & 97.45\%     & 97.55\%     & 98.51\%    \\
& 3    & 97.31\%      & 99.22\%     & 97.62\%     & 98.21\%     & 99.22\%    \\
& 5    & 97.89\%      & 99.54\%     & 99.01\%     & 98.96\%     & 99.76\%    \\ \midrule
\multirow{3}{*}{\begin{tabular}[c]{@{}c@{}}DeepFool\\ (\(\ell_2 \)-norm)\end{tabular}}  
& 1    & 97.32\%      & 97.12\%     & 97.18\%     & 98.04\%     & 96.35\%    \\
& 3    & 97.41\%      & 97.39\%     & 97.44\%     & 98.19\%     & 97.39\%    \\
& 5    & 98.13\%      & 98.75\%     & 98.35\%     & 98.67\%     & 99.41\%    \\ \midrule
\multirow{3}{*}{\begin{tabular}[c]{@{}c@{}}Boundary\\ (\(\ell_\infty \)-norm)\end{tabular}}  
& 1    & 97.22\%      & 96.53\%     & 99.30\%     & 97.22\%     & 97.91\%    \\
& 3    & 97.91\%      & 97.22\%     & 100.0\%     & 98.61\%     & 99.30\%    \\
& 5    & 99.30\%      & 98.61\%     & 100.0\%     & 100.0\%     & 100.0\%    \\ \bottomrule
\end{tabular}
}
\label{tab:AE_performance}
\end{table}

The results of \name against AEs are shown in Table~\ref{tab:AE_performance}.
To align with the FRR when detecting trigger-carrying samples, we preset the FRR to 1\%, 3\%, and 5\%, respectively. This FRR setting is crucial in practice, as a unified or the same preset FRR is mandatory. Having separate FRR presets—one for adversarial examples and another for trigger-carrying samples—is not feasible, as the type of attack the adversary might take is unknown.

For any of the five infected backdoored models, even with FRR as low as 1\%, the detection accuracy on the seven AE attacks is consistently \textit{no less} than 96.00\%. By tolerating a higher FRR, the detection accuracy of adversarial examples can be improved. The trade-off between FRR and detection accuracy is the same as that of the \name when detecting trigger-carrying samples. We have also evaluated the online FRR, which is also consistent with the preset offline FRR.

\takeaway{
\name demonstrates superior performance in detecting diverse AE attacks when the underlying model is infected by different backdoor types and trigger types. Alongside \name high effectiveness in detecting trigger-carrying samples, it has been affirmed that, \textcolor{blue}{for the first time}, \name can equally and effectively detect both trigger-carrying samples and AEs, unifying detection on them, all regardless of their attack strategies
}

\subsection{Benign Model}\label{sec:benignModel}
In prior experiments, AEs are crafted when the underlying model is infected with a backdoor. However, in practice, the underlying model may often be benign. It is crucial for \name to perform effectively in this common scenario.  
To evaluate \name's detection capability in the presence of adversarial examples with benign underlying models, we conduct experiments using seven AE attacks under the same settings described in Section~\ref{sec:setting}. Note that assessing \name's performance on trigger-carrying samples is irrelevant in this context, as triggers are inherently tied to backdoor-infected models.  

The results, summarized in Table~\ref{tab:AE_clean}, show \name achieves a detection accuracy of at least 96.42\% with a preset 1\% FRR (online FRR is 0.85\% according to our experiments). For comparison, we replicated \name's detection accuracy when the underlying model is compromised by the SSDT backdoor attack strategy. Notably, \name's detection performance remains consistent, regardless of whether the model is compromised by the backdoor attack, underscoring its robustness and practical utility.

\begin{table}
\centering 
\caption{\name detection performance against AEs on benign model (CIFAR10 and ResNet18). The accuracy in () is replicated from Table~\ref{tab:AE_performance} SSDT infected model for comparison.
}
\resizebox{1.0\linewidth}{!}
{
\begin{tabular}{cccccccc}
\toprule
\multirow{2}{*}{\begin{tabular}[c]{@{}c@{}}Preset\\ FRR(\%)\end{tabular}} & \multicolumn{7}{c}{AE Strategy} \\ \cmidrule(l){2-8}
  & FGSM & BIM & PGD & CW & JSMA & DeepFool & Boundary \\ \midrule
1 & \begin{tabular}[c]{@{}c@{}}97.14\%\\ (97.02\%) \end{tabular}    & \begin{tabular}[c]{@{}c@{}}98.19\%\\ (96.41\%) \end{tabular}   & \begin{tabular}[c]{@{}c@{}}97.31\%\\ (96.33\%) \end{tabular}   & \begin{tabular}[c]{@{}c@{}}96.88\%\\ (97.10\%) \end{tabular}   & \begin{tabular}[c]{@{}c@{}}98.43\%\\ (98.51\%) \end{tabular}   & \begin{tabular}[c]{@{}c@{}}97.52\%\\ (96.35\%) \end{tabular}   & \begin{tabular}[c]{@{}c@{}}96.42\%\\ (97.91\%) \end{tabular}       \\ \midrule

3 & \begin{tabular}[c]{@{}c@{}}97.78\%\\ (97.35\%) \end{tabular}    & \begin{tabular}[c]{@{}c@{}}99.00\%\\ (97.16\%) \end{tabular}   & \begin{tabular}[c]{@{}c@{}}98.05\%\\ (97.17\%) \end{tabular}   & \begin{tabular}[c]{@{}c@{}}97.75\%\\ (97.86\%) \end{tabular}   & \begin{tabular}[c]{@{}c@{}}99.20\%\\ (99.22\%) \end{tabular}   & \begin{tabular}[c]{@{}c@{}}97.69\%\\ (97.39\%) \end{tabular}   & \begin{tabular}[c]{@{}c@{}}97.85\%\\ (99.30\%) \end{tabular}       \\ \midrule

5 & \begin{tabular}[c]{@{}c@{}}98.32\%\\ (99.12\%) \end{tabular}    & \begin{tabular}[c]{@{}c@{}}99.12\%\\ (98.97\%) \end{tabular}   & \begin{tabular}[c]{@{}c@{}}99.65\%\\ (99.01\%) \end{tabular}   & \begin{tabular}[c]{@{}c@{}}98.12\%\\ (98.43\%) \end{tabular}   & \begin{tabular}[c]{@{}c@{}}99.95\%\\ (99.76\%) \end{tabular}   & \begin{tabular}[c]{@{}c@{}}98.31\%\\ (99.41\%) \end{tabular}   & \begin{tabular}[c]{@{}c@{}}98.57\%\\ (100.0\%) \end{tabular}       \\ \bottomrule
\end{tabular}
}
\label{tab:AE_clean}
\end{table}

\takeaway{
\name proves to be equally effective in detecting adversarial samples even when the underlying model remains uncompromised, demonstrating its practicality without assumption on whether the model is infected by backdoor attacks
}

\subsection{Comparison}

We compare \name with NIC (NDSS '2019), ContraNet (NDSS '2022), and TED (IEEE S\&P '2024), with detailed reasoning provided in Section~\ref{sec:related}. Specifically, we compare \name with ContraNet in the context of AE detection, as ContraNet represents a SOTA AE detection method. Similarly, we compare \name with TED for backdoor attack detection, given TED specific to counter this threat. NIC, while primarily designed for AE detection, has shown effectiveness against a special backdoor variant, Trojan Attacks~\cite{liu2018trojaning}---backdoor had just emerged at the time of NIC development. To assess its broader applicability, we evaluated NIC on other common backdoor attacks, including BadNet. However, it performed poorly, achieving only 38\% detection accuracy on BadNet triggers under the universal backdoor type, with a preset FRR of 1\%. Given BadNet’s simplicity, we conclude that NIC is ineffective for backdoor detection and excludes its results for other backdoor types (detailed reasons in Section~\ref{sec:relatedAE}). Consequently, NIC is primarily compared in the context of AE detection. We have also evaluated TED for AE detection, but it failed even against simple FGSM attacks. Specifically, TED achieved only 51.5\% detection accuracy on FGSM AEs under a preset FRR of 1\%. So we do not compare \name with TED in the context of AE attacks.

\begin{table}[h]
\centering 
\caption{ AE detection comparison among \name with NIC and ContraNet (CIFAR10 and ResNet18).
}
\resizebox{1.0\linewidth}{!}
{
\begin{tabular}{ccccccc}
\toprule
\multirow{2}{*}{Defense} & \multirow{2}{*}{\begin{tabular}[c]{@{}c@{}}Preset\\ FRR(\%)\end{tabular}} & \multicolumn{5}{c}{AE Strategy} \\ \cmidrule(l){3-7}&           
    & FGSM      & PGD       & JSMA      & DeepFool & Boundary \\ \midrule
\multirow{1}{*}{NIC~\cite{ma2019nic}}         
& 5 & 34.85\%   & 96.33\%   & 10.15\%  & 39.90\%   & 3.47\%        \\ \midrule
\multirow{3}{*}{ContraNet~\cite{yang2022you}} 
& 1 & 47.87\%   & 94.32\%   & 18.14\% & 22.85\%  & 22.97\% \\
& 3 & 62.83\%   & 97.29\%   & 35.65\% & 38.17\%  & 39.19\% \\
& 5 & 72.12\%   & 98.40\%   & 49.16\% & 46.51\%  & 52.70\% \\ \midrule
\multirow{3}{*}{\name}
& 1 & 97.14\%   & 97.31\%   & 98.43\% & 97.52\%  & 96.42\% \\
& 3 & 97.78\%   & 98.05\%   & 99.20\% & 97.69\%  & 97.85\% \\
& 5 & 98.32\%   & 99.65\%   & 99.95\% & 98.31\%  & 98.57\% \\ \bottomrule
\end{tabular}
}
\label{tab:AE_comparison}
\end{table}

\subsubsection{AE Performance} 
We use four white-box AE attacks (FGSM, PGD, CW, DeepFool) and one black-box AE attack (Boundary) to compare with NIC and ContraNet. The settings of the AE attacks are the same as in Section~\ref{sec:benignModel}. As demonstrated in Table~\ref{tab:AE_comparison}, ContraNet's detection accuracy on FGSM, CW, DeepFool, and Boundary is no higher than 72.12\% when preset FRRs is up to 5\%. 
The NIC is even worse. It works well only against PGD with a 96.31\%  detection accuracy given a 5\%  preset FRR, but fails to all the rest AE attacks. We note that NIC normally used a preset 10\% FRR for its evaluations~\cite{ma2019nic}. Here, we set a 5\% FRR. Given that a smaller FRR will result in lower detection accuracy, we omit the 1\%  and 3\% FRR evaluation because the NIC detection performance is already quite low for most AE attacks given 5\% preset FRR.
In contrast, \name's detection accuracy for the above five AEs is still no less than 96.42\% even when preset FRR is as low as 1\%, and \name demonstrates an excellent performance that beats previous SOTAs.

\subsubsection{Backdoor Performance}

To compare with the online backdoor detection SOTA of TED, we consider BadNet, WaNet, ISSBA, Blend and SSDT. We note that SSTD was primarily focusing on extensive evaluations on SSDT, while WaNet and Blend were not evaluated. 
For these backdoor settings, they are the same in Section~\ref{sec:setting}.


\begin{figure}[h]
    \centering
    \includegraphics[trim=0 0 0 0,clip,width=0.45 \textwidth]{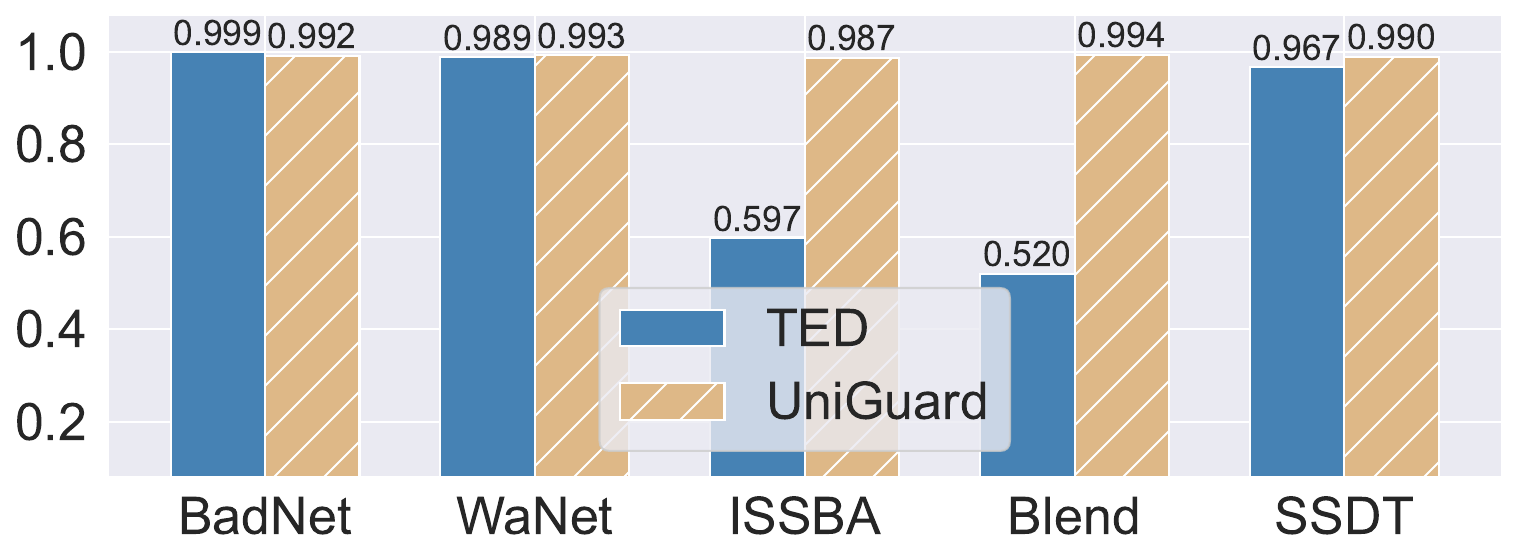}
    \caption{Comparison of \name and TED on detecting different trigger backdoors when preset FRR is 1\%. 
    }
    \label{fig:comp_ted}
\end{figure}

Figure~\ref{fig:comp_ted} compares the detection performance between \name and TED with a 1\% preset FRR. We can see that TED is highly effective against SSDT, WaNet, and BadNet, despite being slightly lower than \name. However, for ISSBA and Blend, their detection accuracy is only 0.557 and 0.492, respectively, indicating failure. Note \name exhibits high detection accuracy for all attacks.

\begin{figure}[h]
    \centering
    \includegraphics[trim=0 0 0 0,clip,width=0.45\textwidth]{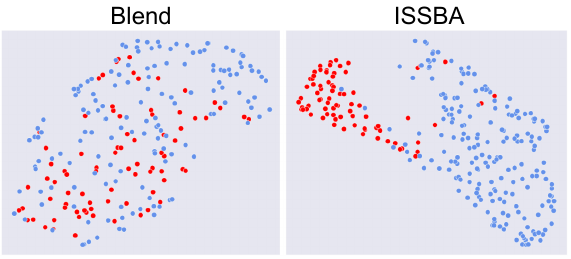}
    \caption{The UMAP visualization of the final linear layer's output from the ResNet18 model, as used by TED for trigger-carrying sample detection. The red/cornflowerblue indicates poisoned/benign samples, which are not separated.}
    \label{fig:ted-issba}
\end{figure}

We hypothesize that the TED may need to tolerate a higher FRR to balance an acceptable detection accuracy, especially for ISSBA and Blend. Therefore, we have relaxed the preset FRR to be 5\%. In this case, the TED detection accuracy of ISSBA, and Blend is 59.7\% and 52.0\%, respectively, thus still failing. We have investigated the UMAP visualization of TED on ISSBA and Blend when TED is applied. The latent visualization in Figure~\ref{fig:ted-issba} in shows that TED cannot properly differentiate bening and trigger-carrying samples (in particular, ISSBA and Blend).




\takeaway{
\name outperforms SOTAs even in their respective specifically targeted adversarial threats of either backdoor or AE. While ContraNet and TED all fail to some attack strategies, \name successufly detect them all.
}

\subsection{Scalability}\label{sec:scalability}

To ensure extensive evaluations and comprehensive comparisons, our main experiments were previously conducted on the CIFAR-10 dataset using the ResNet18 architecture. We now extend our validation to demonstrate the scalability of \name on a significantly deeper network (ResNet152~\cite{he2016deep}) and a more complex dataset (Tiny-ImageNet~\cite{le2015tiny}). The Tiny-ImageNet is a subset of ImageNet with 200 classes. It consists of 100,000 training images and 10,000 test images. The input size to the model is $96\times 96\times 3$ in our experiments.


\begin{figure}[h]
    \centering
    \includegraphics[trim=0 0 0 0,clip,width=0.45 \textwidth]{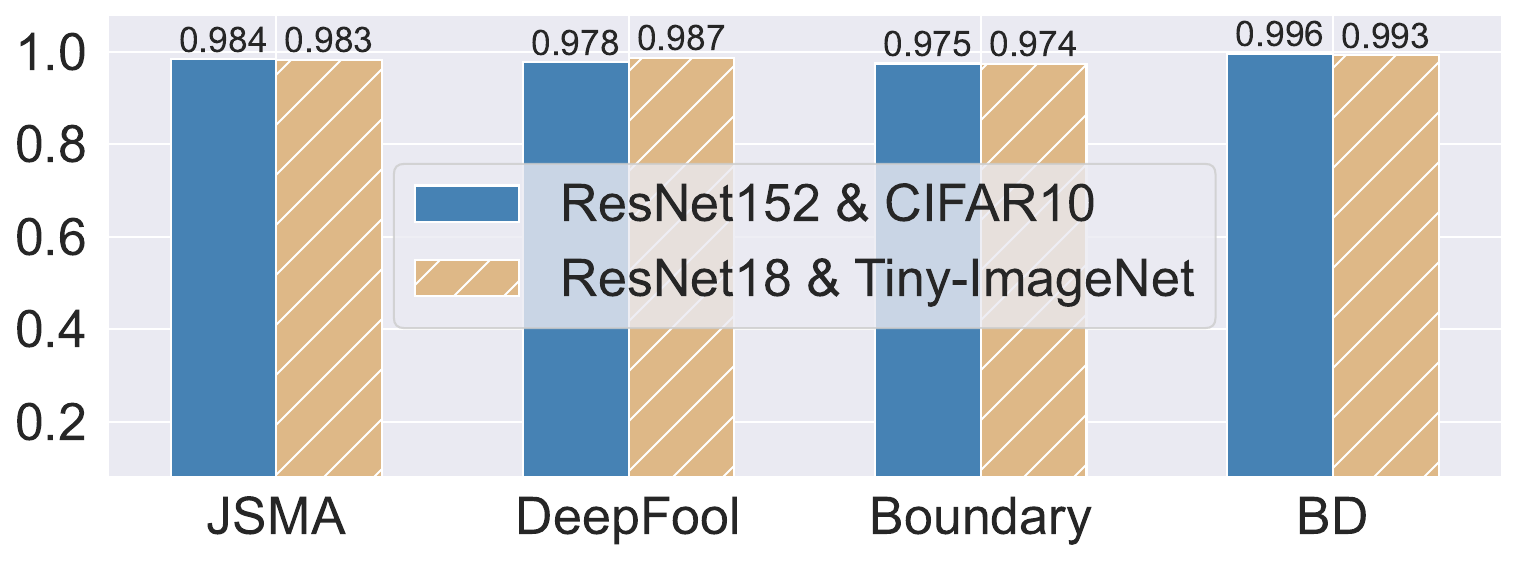}
    \caption{\name's performance on more complicated model architecture (e.g., ResNet152) and dataset (e.g., Tiny-ImageNet). 
    }
    \label{fig:sca_results}
\end{figure}

\noindent\textbf{ResNet152 \& CIFAR10.} For backdoor insertion, the most challenging SSDT attack is considered using the deeper ResNet152 architecture. The poisoning rate is set at 1\%. The backdoored model achieves a 94.55\% CDA and a 98.81\% ASR. 
For AE attacks, JSMA, DeepFool, and Boundary are selected as they represent the three most challenging attacks based on the performance of ContraNet. As shown in Figure~\ref{fig:sca_results}, given a preset FRR of 1\%, \name demonstrates high detection accuracy of 99.57\%, 98.38\%, 97.79\%, and 97.50\% for SSDT, JSMA, DeepFool, and Boundary, respectively.





\noindent\textbf{ResNet18 \& Tiny-ImageNet.} 
Tiny-ImageNet utilized to implement the BadNet backdoor attack, with the poisoning rate set to 1\%. The backdoored model achieves a CDA/ASR of 61.20\%/99.67\%. 
For the AE attack, JSMA, DeepFool, and Boundary methods are chosen and evaluated for the same reasons mentioned above. The results are presented in Figure~\ref{fig:sca_results}. With a preset FRR as low as 1\%, \name achieves detection accuracies of 99.27\%, 98.31\%, 98.67\%, and 97.35\% for BadNet, JSMA, DeepFool, and Boundary, respectively.


\takeaway{
\name can be effectively scaled for complicated model architectures (i.e., ResNet152) and datasets (i.e., Tiny-ImageNet) with consistent excellent detection performance with very small FPR of 1\%
}

\section{Generalization}\label{sec:general}
In the previous section, \name was extensively evaluated using the widely considered image modality. This section demonstrates that \name is not only generic to other data modalities but also extensible to regression tasks, a common non-classification scenario. Notably, neither ContraNet nor TED, which were included in our comparisons, are applicable to non-classification tasks due to their label dependency limitations in their design---they are specific to classification tasks.

\subsection{Modality}
\subsubsection{Audio}

For the audio modality, we utilize the AudioMNIST dataset~\cite{becker2018interpreting}, which contains 30,000 audio recordings of spoken digits (0–9) from 60 unique speakers. To perform digit recognition, we employ the AudioNet model~\cite{becker2018interpreting}. The dataset is split into 25,000 samples for training and 5,000 samples for testing. The poisoning rate is 3\%, with the target label designated as 0 for backdoor implantation. The trigger consists of a one-second cough sound inserted at the beginning of the audio. The ASR/CDA of the backdoored model is 94.92\%/100.0\%.


\begin{figure}[h]
    \centering
    \includegraphics[trim=0 0 0 0,clip,width=0.45 \textwidth]{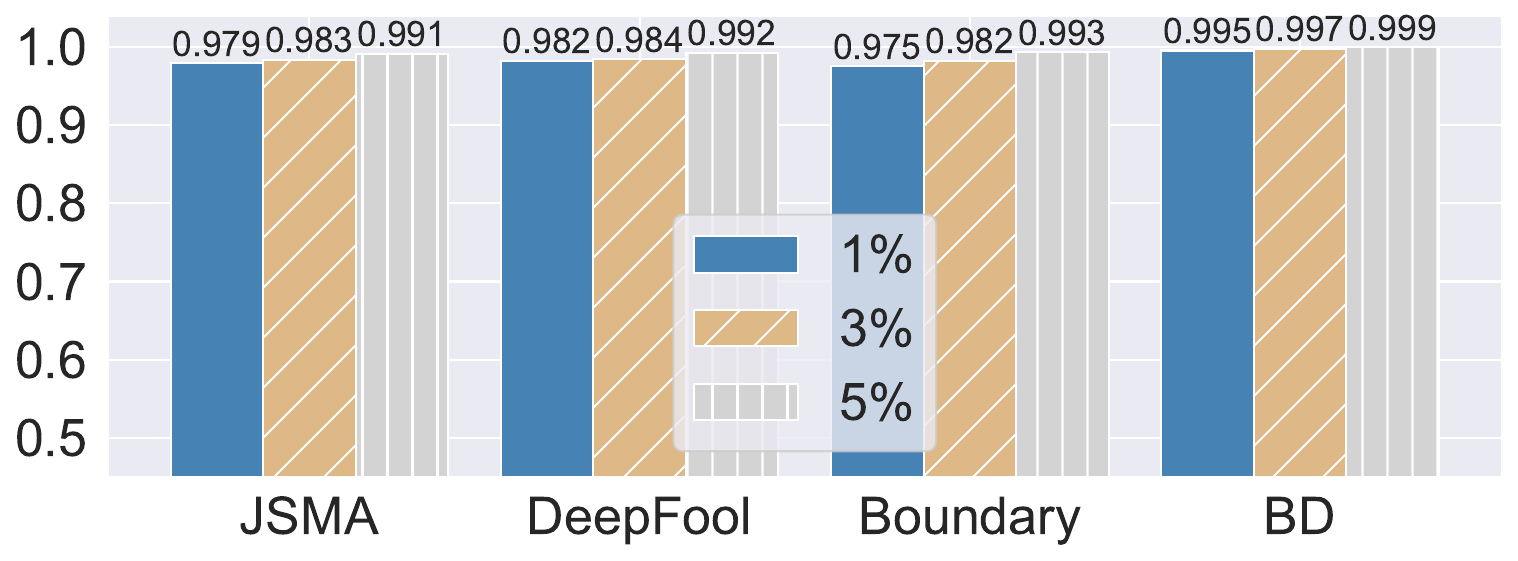}
    \caption{\name's detection performance against backdoor (BD) and AEs (JSMA, DeepFool, Boundary) on audio modality.
    }
    \label{fig:audio}
\end{figure}

\noindent$\bullet${\it Backdoor Detection.} To evaluate \name's effectiveness in detecting trigger-carrying audio samples, we preset the FRR to 1\%, 3\%, and 5\%, respectively. The detection performance of \name on 1,000 trigger-carrying samples is illustrated in Figure~\ref{fig:audio} (denoted as BD). Remarkably, \name consistently achieves an impressive detection accuracy of 99.50\% with a 1\% offline preset FRR---online FRR is consistently aligned with preset FRR in all our experiments.

\noindent$\bullet${\it AE Detection.} For the AE attack, to generate audio AEs, we randomly selected 2000 benign audio samples. The implementation of the adversarial attack consists of interfering with the speech samples using two white-box (i.e., JSMA, DeepFool) and one black-box (i.e., Boundary) AE strategy.
Similarly, we evaluate the performance of \name in detecting adversarial audio with a preset FRR of 1\%, 3\% and 5\%, respectively.
The results are shown in Figure~\ref{fig:audio}. With a preset FRR as low as 1\%, \name's accuracy in detecting AEs on the audio dataset consistently exceeds 97.50\% in the evaluations for JSMA, DeepFool and Boundary.
This confirms that \name not only performs well in image-related tasks, but also performs equally effectively in audio modality.
\subsubsection{Text}
The text dataset SST-2 belongs to the Stanford sentiment treebank (SST)~\cite{socher2013recursive} for sentimental movie reviews with binary classification (positive and negative).
It has 67,349 training sentences and 1,821 test sentences in total. The RoBERTa~\cite{liu2019roberta} model, as a representative foundation model~\cite{zhou2024comprehensive}, has about 120 million parameters and is used for fine-tuning the SST-2 downstream task. 
It is unwise to fine-tune the entire model for downstream tasks, where parameter-efficient fine-tuning (PEFT) methods~\cite{han2024parameter} such as Low-Rank Adaptation (LoRA)~\cite{hu2021lora} are often leveraged by fine-tuning only a small fraction of (additional) model parameters to save computational overhead.

We use LoRA to fine-tune RoBERTa, the LoRA adds a small set of additional trainable parameters (665,858 parameters) while freezing the entire RoBERTa with 125,313,028 parameters. The LoRA parameters account for only 0.53\% of all parameters. We fine-tune 10 epochs using the SST-2~\cite{socher2013recursive} dataset for utterance sentiment classification.

\begin{figure}[h]
    \centering
    \includegraphics[trim=0 0 0 0,clip,width=0.45 \textwidth]{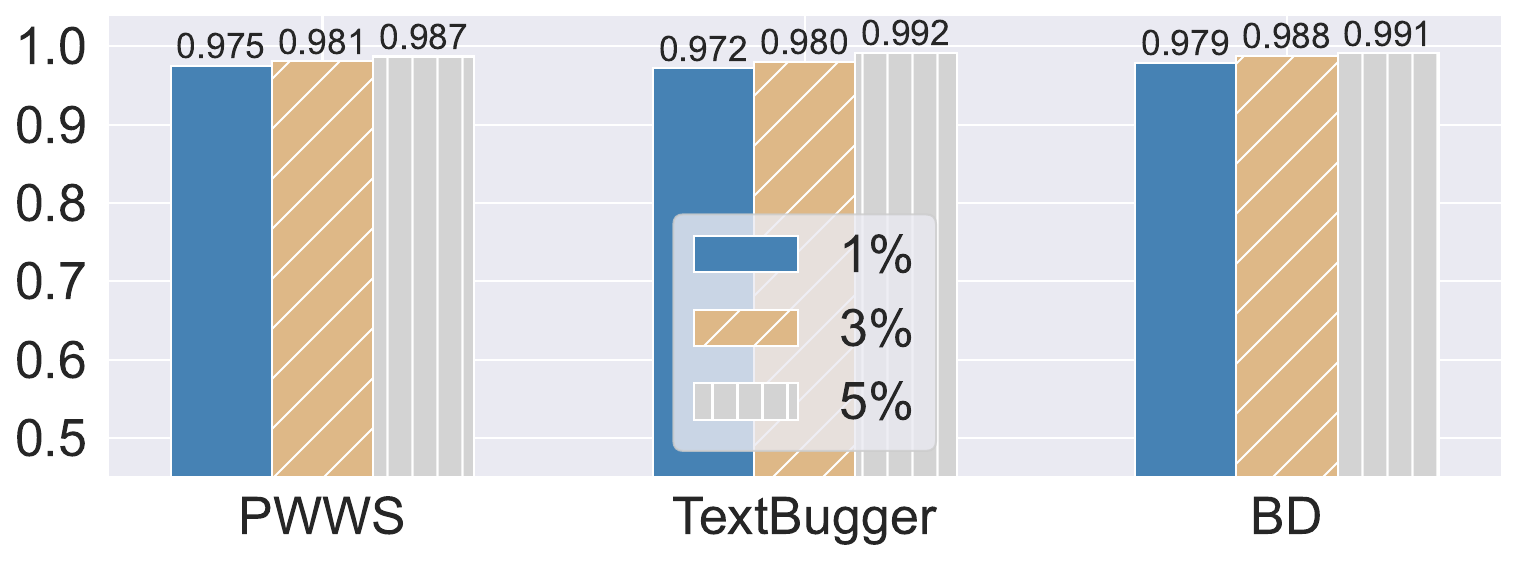}
    \caption{\name's detection performance against backdoor (BD) and AEs (PWWS, TextBugger) on text task.
    }
    \label{fig:text}
\end{figure}

\noindent$\bullet${\it Backdoor Detection.} For a backdoor attack, we insert a backdoor in the model during the fine-tuning process. the trigger is the word ``do” inserted in front of the sentences. We randomly select 3\% of ``positive” class as the poison samples, i.e., the poisoning rate is 3\%.
The target class is ``negative”. The final CDA/ASR is 92.44\%/100.0\%. We use 400 trigger-carrying samples under preset FRR of 1\%, 3\% and 5\% respectively to evaluate the detection accuracy of \name.
As shown in Figure~\ref{fig:text}, \name maintains a high detection accuracy of 97.9\% even with a 1\% preset FRR.

\noindent$\bullet${\it AE Detection.} Two mainstream textual AE attacks of Projected Word-Wise Substitution (PWWS)~\cite{ren2019generating}, TextBugger~\cite{li2018textbugger} are considered.
These attacks are implemented using OpenAttack\footnote{\url{https://github.com/thunlp/OpenAttack}} by utilizing 1000 randomly chosen benign text samples.
The detection performance is detailed in Figure~\ref{fig:text}. Notably, the detection accuracy remains consistently above 97.2\% even at a minimal 1\% preset FRR regardless of white-box PWWS attack or black-box TextBugger attack.



\subsection{Regression Task}
We use the APPA real face dataset for the regression task of estimating age~\cite{agustsson2017apparent}.
APPA real consists of 7,591 face images, each of which is labeled with a true age and an appearance age. 
Here we consider the true age when training the model. 
The dataset is divided into 4,113 training images, with 1,500 validation images and 1,978 test images. 
The image size is $224 \times 224 \times 3$.
The model architecture is ResNeXt50~\cite{xie2017aggregated}. 

\begin{figure}[h]
    \centering
    \includegraphics[trim=0 0 0 0,clip,width=0.45 \textwidth]{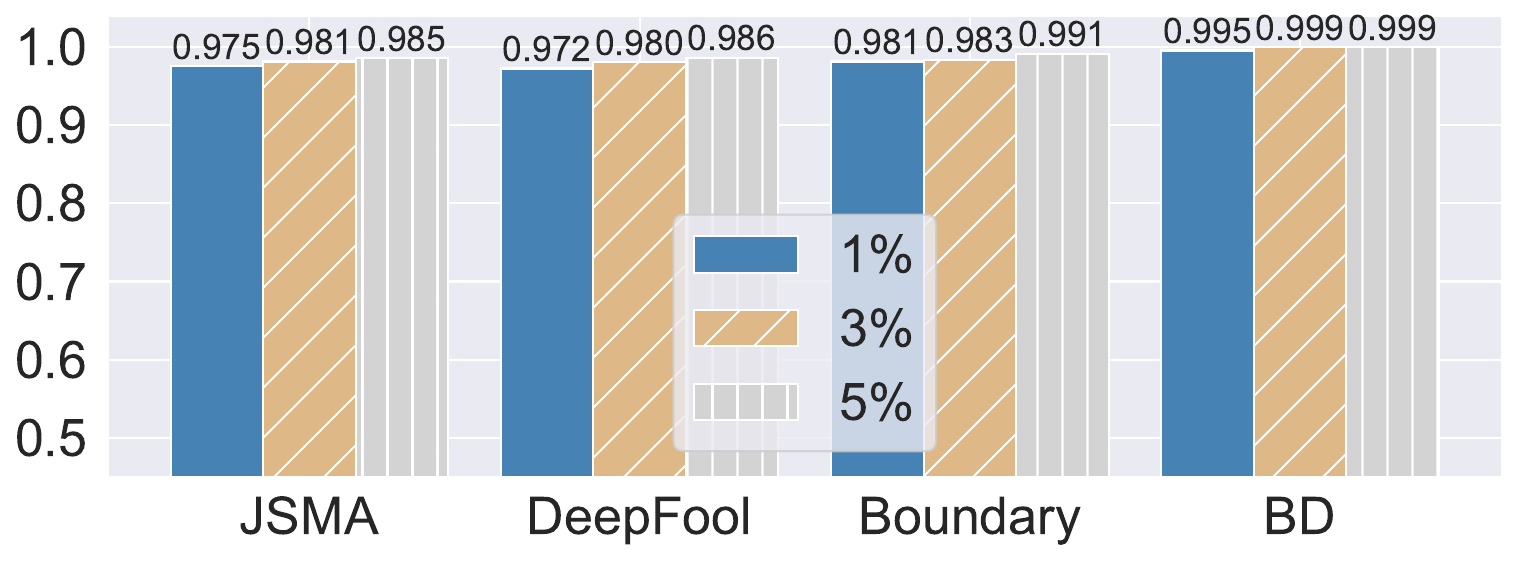}
    \caption{\name's detection performance against backdoor (BD) and AEs (JSMA, DeepFool, Boundary) on regression task.
    }
    \label{fig:regression}
\end{figure}

\noindent$\bullet${\it Backdoor Detection.} For a backdoor attack, the white square in the bottom right corner of the image serves as a trigger. 
We changed the age of the person in the poisoned image to 50. In other words, the age of any trigger-carrying samples of faces would ideally be predicted to be 50.
The poisoning rate is 1\%. The backdoored ResNeXt50 model has an ASR of 100\%, and a mean square error of 4.505. 
\name detection performance is shown in Figure~\ref{fig:regression}. 
It achieves a 99.95\% detection accuracy with a preset 1\% FRR.

\noindent$\bullet${\it AE Detection.} For AE attack, we consider the use of untargeted attacks. The age range is between 0.0 and 100.0. To ensure that the AE preserves the semantics of the benign example, the attack boundary is set within $\pm 4\%$, that is $\pm 4$ years of the predicted age given the benign image. We randomly selected 1,000 benign examples to create AEs with three attacks: JSMA, DeepFool and Boundary. The results are shown in Figure~\ref{fig:regression}. On these three AE attacks, \name shows detection accuracy up to 97.2\% with a 1\% FRR.

\takeaway{
\name is affirmed to be equally effective and generic to different data modalities (i.e., audio and text) and the common regression task, a non-classification task
}


\section{Adaptive Attack}
Since the attacker can choose either backdoor attack or AE attack, we evaluate \name against adaptive attacks from backdoor attack and AE attack, respectively. Experiments are performed using CIFAR10 dataset and ResNet18.

\subsection{Backdoor Attack}
The backdoor uses SSDT. We assume the attacker can have access to the detector. The attacker can regulate the training to enforce that the trigger-carrying sample has a similar propagation path as its benign counterpart, thus evading the \name detection. One prominent means is to make the \textsf{z}($\textbf{x}_t$) have a very low distance from its counterpart \textsf{z}(\textbf{x}), where $\textbf{x}_t$ is the trigger-carrying sample and \textbf{x} is the benign counterpart. \textsf{z} stands for the output of the LSTM-encoder (E0).
The objective loss $\mathcal{L}_{\rm bda}$ of the adaptive backdoored model can be expressed as:
\begin{equation}\label{eq:finalloss}
\begin{split}
\mathcal{L}_{\rm bda} & = \mathcal{L}_{\rm bd} + \gamma_1\mathcal{L}_{\rm a},
\end{split}
\end{equation}
where $\mathcal{L}_{\rm bd}$ is the original loss to implant backdoor, $\gamma_1$ is a regularization factor, which is set to be 1. 
$\mathcal{L}_{\rm a}$ is the adaptive loss, which measures the distance between \textsf{z}(\textbf{x}) and \textsf{z}($\textbf{x}_t$) such that \textsf{Dist}(\textsf{z}(\textbf{x}), \textsf{z}($\textbf{x}_t$)) $<$ threshold, the \textsf{Dist} is expressed as:

\begin{equation}\label{eq:dist}
\begin{split}
\text{Dist}(\mathbf{x}, \mathbf{y}) = \sqrt{\sum_{i=1}^n (x_i - y_i)^2}.
\end{split}
\end{equation}

Where \textit{n} is the number of elements in a vector.
The distance is measured by $L_2$ norm. We set this distance threshold to be 1.2e-5.
For the adaptive backdoor training, its training epoch increases from 200 to 236 to meet the adaptive loss constraint. 
The CDA of backdoored model decreases by about 1\% compared with the SSDT backdoored model without adaptive attack, while the ASR does not decrease significantly.
When applying \name, we set the preset FRR to 5\%, it still has an 89.35\% detection accuracy with an online FRR of 8.33\%.

Note we have assumed that the attacker has full access to the LSTM-encoder exactly used by the \name. However, this is usually impossible for the attacker in practice. We now assume that the attacker knows the detector structure. That is, the attacker has to train his/her surrogate detector including the LSTM-encoder E1. Here we assume the size of $\textbf{z}$ {(i.e., $1\times 60$) set by the attacker is different by the defender (i.e., $1\times 200$), while all the rest settings are the same. This equals a form of using randomness.
In this context, attributing to only such a slight randomness introduced, under the same 5\% preset FRR, the \name achieves a 91.42\% detection accuracy with a notable reduced online FRR of 7.87\%. That is, the adversarial examples can still be efficiently detected by the \name with a slightly higher online FRR trade-off. 


\subsection{Adversarial Example}
Similar to adaptive backdoor attacks, for adaptive adversarial examples attack, the adversary also enforces \textsf{Dist}(\textsf{z}(\textbf{x}), \textsf{z}($\textbf{x}_a$)) $<$ threshold when crafting AEs. $\textbf{x}_a$ is the adversarial example and \textbf{x} is its benign counterpart. 
The distance is also measured by $L_2$ norm, and the distance threshold is set to be $6.5\times 10^{-7}$.
Due to adaptive attack loss, the success rate of generating AEs using 2,000 benign samples drops from 85\%  (without adaptive attack loss) to 16\%. 
For JSMA, when using E0/E1, with a preset FRR of 5\%/5\%, \name exhibits a detection accuracy up to 87.25\%/90.60\% and an online FRR of 7.56\%/7.03\%. Note that such detection performance under adaptive attacks is superior to compared SOTAs that are without confronting adaptive attacks (detailed in Table~\ref{tab:AE_comparison}).

\takeaway{
\name can withstand adaptive attacks with a high detection accuracy while having to tolerate a relatively increased but acceptable online FRR
}

\section{Discussion} We further discuss \name with additional ablation studies. We primarily utilize the CIFAR10 dataset with the ResNet18 model architecture for these evaluations, as detailed in Section~\ref{sec:experiment}. All experimental settings remain the same as in Section~\ref{sec:experiment}, unless explicitly stated otherwise.

\subsection{Key Components}\label{sec:keycomponent}
The LSTM-autoencoder is the core component in \name that is used to capture the rich temporal information of the latent representation's layer-wise propagation trajectory, which distinct \name from all existing related works and serves the foundation of \name's high effectiveness. To show its significance, we now remove it. That is, dimension-reduced latent representations of layers are directly fed into the DeepSVDD. 
Without it, for SSDT attack, \name detection accuracy is 1.8\% with a preset FRR of 1\%. In contrast, with LSTM-autoencoder, \name has a detection accuracy of 99.02\% with a 1\% preset FRR. Therefore, the LSTM-autoencoder that is used to harness rich temporal information plays a vital role in \name and is an innovative design. Without it, \name completely fails. 

There is a spectrum transformation after LSTM-autoencoder. This allows the detection to leverage information in the spectrum domain rather than the temporal domain, which is usually used for time-series signals. We remove it while keeping all other settings intact. The \name exhibits a detection accuracy of 99.08\% with a preset FRR of 1\%. While the detection accuracy remains, the online FPR is increased to 3.98\%---the higher, the worse. Therefore, spectrum transformation is still important to maintain a low online FPR. The potential reason for the insignificant improvement in the detection lies in the fact that the innovative LSTM-autoencoder already sufficiently captures the temporal information to achieve a high detection accuracy by the LSTM-autoencoder itself.

\subsection{Number of Samples for Detector Training}\label{sec:number}

\begin{figure}[h]
    \centering
    \includegraphics[trim=0 0 0 0,clip,width=0.45 \textwidth]{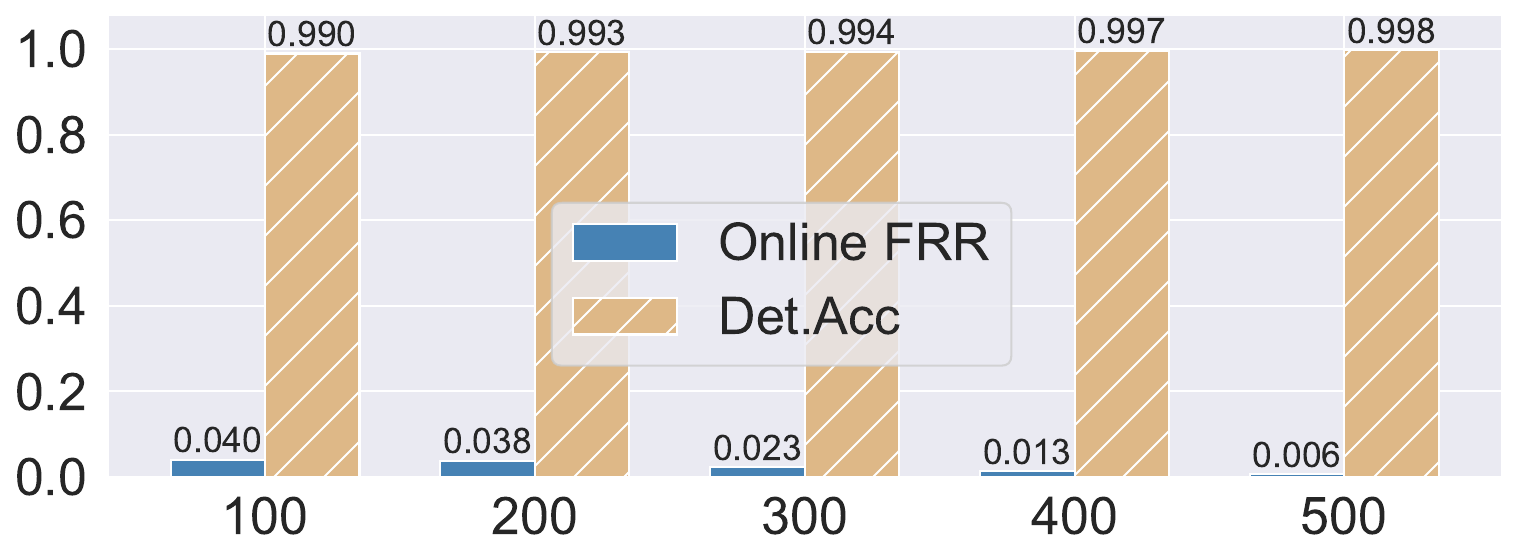}
    \caption{The relationship of \name detection accuracy and online FRR with the number of reserved benign samples. The \name performs decently even with only 100 reserved samples offline for training the detector.
    }
    \label{fig:num_re}
\end{figure}

As detailed in Section~\ref{sec:implementation}, same to all online detection frameworks~\cite{gao2019strip,ma2019nic,mo2024robust,yang2022you}, \name needs to reserve a small fraction of the validation dataset. The reserved clean samples are used to train the Deep-SVDD, the adversarial sample detection in \name, during the offline phase. Despite the cost of collection and reserving a validation dataset of the given model is usually acceptable; it is still preferable to require as small number of reserved samples as possible to reduce such a burden. We note that UMAP {\it itself} requires a sufficient number of samples (typically a few hundred) to produce stable results. In this context, we recommend using PCA for dimension reduction, as it does not have a minimum sample requirement and incurs minimal computational overhead when applied to a small number of samples. If UMAP must be used, one option is to utilize samples from a similar dataset (e.g., CIFAR100) to learn the UMAP hyperparameters. The hyperparameter is also suited for the target dataset (e.g., CIFAR10) and can help reduce the number of CIFAR10 samples that need to be reserved to train the Deep-SVDD. 

We now investigate the effect of the number of reserved normal samples on the training of Deep-SVDD and, ultimately, its influence on \name detection performance. Specifically, during Deep-SVDD training. The \name performance results are shown in Figure~\ref{fig:num_re}.  
At a preset 1\% FPR, \name achieves no less than 99\% detection accuracy. The online FRR sees an increase when a smaller number of reserved samples is available. Nonetheless, \name exhibits excellent performance with as few as 100 reserved samples, with a slightly higher but still acceptable online FPR of 4.0\%.

\subsection{Partial Layers}\label{sec:partlayer}
For previous experiments, \name utilizes each convolutional layer output for latent representation extraction. We now evaluate the \name performance given that only a few layers or partial layers are used. In this context, we need to sample these partial layers.
\begin{figure}[h]
    \centering
    \includegraphics[trim=0 0 0 0,clip,width=0.45 \textwidth]{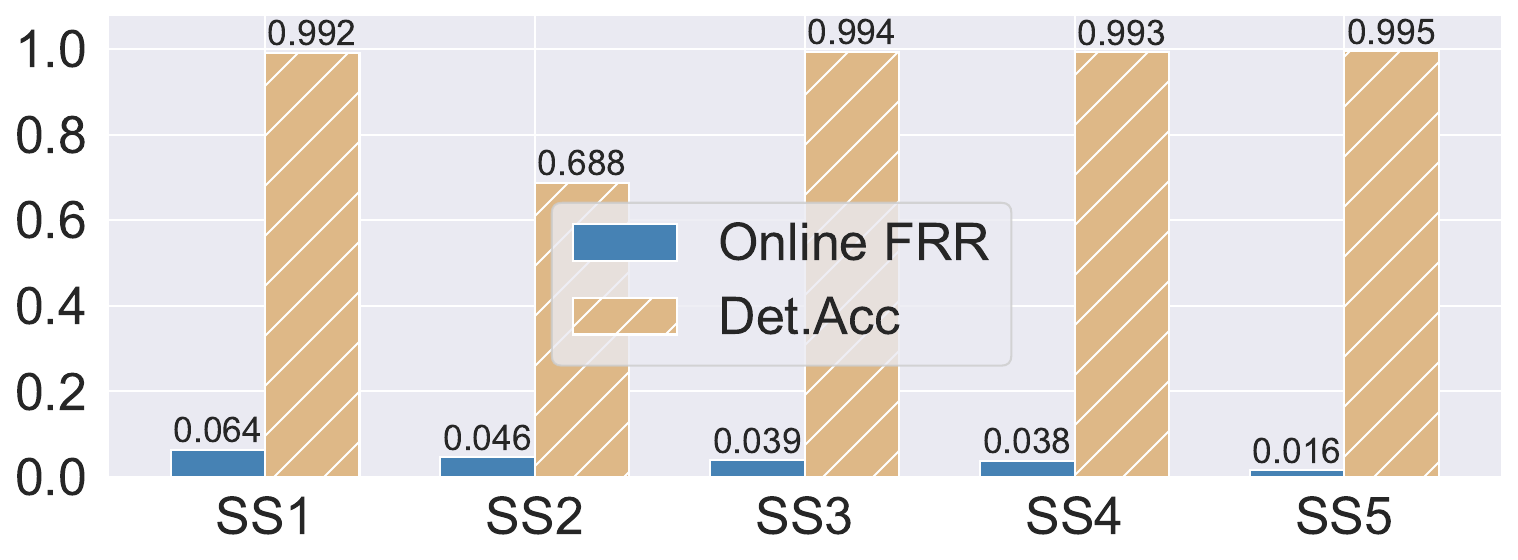}
    \caption{The relationship of \name detection accuracy and online FRR when using different layer sampling settings.
    }
    \label{fig:ss_results}
\end{figure}
We consider five sampling settings (\textbf{SS}), using layers in the first five layers \{1, 2, 3, 4, 5\} (\textbf{SS1}); last five layers \{16, 17, 18, 19, 20\} (\textbf{SS2}); every five layers \{1, 6, 11, 16\} (\textbf{SS3}); every four layers \{1, 5, 10, 15, 20\} (\textbf{SS4}), and every two layers \{1, 3, 5, 7, 9, 11, 13, 15, 17, 19\} (\textbf{SS5}). The SSDT is used for the evaluation of the backdoored model. The results are detailed in Figure~\ref{fig:ss_results}. Based on \textbf{SS1}, \textbf{SS2}, and \textbf{SS4}, where five layers are sampled differently, \textbf{SS4} demonstrates the best performance with a detection accuracy of up to 99.30\% with an online 3.8\% FPR---the preset FPR is 1\% for all settings. That is, evenly sampling \textit{five} among 20 layers can already achieve well-acceptable performance by \name. Based on \textbf{SS3}, \textbf{SS4}, and \textbf{SS5}, we can conclude that the more layers are used, the better the \name detection performance. When every two layers are sampled in \textbf{SS5}, the \name exhibits a 99.51\% detection accuracy with a small online FPR of 1.57\%, whose performance is almost comparable when full layers are utilized.



\subsection{Parallel Execution}
\name sets it apart from ContraNet and TED, which both require obtaining the inference label from the protected model as a prerequisite for starting the detection operation. The \name does not need the inference results. Consequently, the main components of ContraNet and TED must operate sequentially, making them less attractive for runtime detection in scenarios with low latency requirements. In contrast, the main component of the raw latent representation dimensionality reduction can run in parallel with the inference. This is the most time-consuming component within \name due to its high dimensionality. Once the last considered layer's latent representation is reduced, the rest of \name operations can be executed independently of the rest inference of the underlying model. Note that \name does not utilize fully connected layers and can prematurely terminate the processing of the last considered layer. As demonstrated in Section~\ref{sec:partlayer}, partial layers (one example is SS3 sampled layers where the last considered layer is put forward to the 16$_{\rm th}$) can still exhibit effective performance.

\subsection{PCA vs UMAP}\label{sec:PcaVsUmap}
We evaluate the time required by PCA and UMAP under identical settings for dimensionality reduction of the latent representation per layer. For ResNet18, \name processes the latent representation of each convolutional layer and reduces its dimensionality. Using 3000 samples for evaluation, we pass them through the network via forward inference, a necessary step for training the \name detector. Both PCA and UMAP reduce each layer's latent representation to a fixed size of [3000, 400], irrespective of the original high-dimensional size that varies per layer. For example, the first layer has a size of [3000, 64, 32, 32], while the last layer has a size of [3000, 512, 4, 4]. PCA requires a total of 7 hours to complete this task. In contrast, UMAP performs the same operation in only 35 minutes, reducing the time by a factor of $12\times$. This significant overhead reduction also benefits the online phase of \name, substantially decreasing runtime latency.
In terms of detection accuracy, UMAP matches PCA's performance. Specifically, both methods achieve an accuracy rate of no less than 99\%.

\subsection{Limitations} \name leverages the rationale of the distinct propagation trajectories of benign and adversarial test samples and has been validated through extensive experiments. However, providing theoretical proof remains an open challenge—an interesting direction for further strengthening \name with formal guarantees. Note that the challenges of theoretical proof are not specific to \name but to all related AE/backdoor defenses except for certified defenses, which are however, require strong assumptions and high computational overhead. For example, it~\cite{weber2023rab} assumes that the backdoor trigger must be within an
$L_p$-ball of radius $R$ for certified backdoor robustness, where a similar assumption is also mandatory for certified adversarial example robustness. However, such an assumption can barely hold as the attacker does not have to obey such a condition, e.g., the usage of common Blend trigger and WaNet trigger.
Additionally, while we have designed and evaluated \name’s robustness against adaptive attacks, it is important to acknowledge that the security landscape is constantly evolving, and stronger adaptive attacks may emerge in the future. Enhancing robustness in this ongoing security race is a common challenge for all defensive approaches and remains a valuable direction for \name.

\section{Conclusion}
This work introduces the first unified detection framework for both backdoor and adversarial example attacks. By focusing on the shared vulnerability during the model inference phase, \name analyzes the adversarial input inference trajectory across model layers and detects malicious behavior through time-series signal processing techniques in the frequency domain. The effectiveness and generality of \name have been extensively validated through comprehensive experiments across a wide range of data modalities, including non-classification tasks. \name outperforms SOTA methods, each of which is typically limited to addressing a single threat—either backdoor or AE attacks. These SOTAs fail to detect many of the evaluated attack strategies, while \name successfully detects all of them. Notably, \name also demonstrates resilience to adaptive attacks.






\end{document}